\newif\ifanon\anonfalse    %
\ifanon
\documentclass[manuscript, screen, review, anonymous, acmtog]{acmart}
\else
\documentclass[acmtog, nonacm]{acmart}
\fi
\newif\iffull\fulltrue
\newif\ifdraft\draftfalse
\usepackage[compact]{titlesec}
\usepackage{microtype}
\usepackage{amsmath}
\usepackage{xspace}
\usepackage{mathtools}
\usepackage{attrib}
\usepackage{xcolor}
\usepackage{listings}
\usepackage{multirow}

\newcommand\maybecolor[1]{\color{#1}}
\definecolor{addition}{rgb}{0,0.1,0.5}
\definecolor{dkblue}{rgb}{0,0.1,0.5}
\definecolor{dkgreen}{rgb}{0,0.4,0}
\definecolor{dkred}{rgb}{0.6,0,0}
\definecolor{dkpurple}{rgb}{0.7,0,1.0}
\definecolor{purple}{rgb}{0.9,0,1.0}
\definecolor{olive}{rgb}{0.4, 0.4, 0.0}
\definecolor{teal}{rgb}{0.0,0.4,0.4}
\definecolor{azure}{rgb}{0.0, 0.5, 1.0}
\definecolor{gray}{rgb}{0.5, 0.5, 0.5}
\definecolor{dkgrey}{rgb}{0.2, 0.2, 0.2}
\definecolor{lilac}{rgb}{0.70, 0.04, 0.08}

\let\ls\lstinline
\newcommand\function{\ensuremath{\mbox{\sffamily\maybecolor{dkblue}function}}}
\newcommand{\lsf}[1]{\text{\lstinline[language=fstar]|#1|}}
\newcommand{\lsc}[1]{\text{\lstinline[language=C]|#1|}}

\lstdefinelanguage{cddl}{%
  basicstyle=\small,
  morestring=[b]",
  sensitive=true,%
  numbersep=4pt,
  columns=[l]fullflexible,
  texcl=true,
  mathescape=true,
  identifierstyle={\sffamily},
  keywordstyle=[1]{\sffamily\bfseries\maybecolor{dkblue}},
  keywordstyle=[2]{\sffamily\bfseries\maybecolor{dkblue}},
  keywordstyle=[3]{\itshape\maybecolor{dkred}},
  keywordstyle=[4]{\rmfamily\itshape},
  rangeprefix=(*---\ ,
  includerangemarker=false,
  stringstyle=\ttfamily,
  showspaces=false,
  morecomment=[l]{;},
  commentstyle={\itshape\maybecolor{dkred}}
}

\colorlet{punct}{red!60!black}
\definecolor{delim}{RGB}{20,105,176}
\colorlet{numb}{magenta!60!black}

\lstdefinelanguage{json}{
  basicstyle=\small,
  stringstyle=\ttfamily,
    numberstyle=\scriptsize,
    stepnumber=1,
    numbersep=8pt,
    showstringspaces=false,
    literate=
     *{0}{{{\color{numb}0}}}{1}
      {1}{{{\color{numb}1}}}{1}
      {2}{{{\color{numb}2}}}{1}
      {3}{{{\color{numb}3}}}{1}
      {4}{{{\color{numb}4}}}{1}
      {5}{{{\color{numb}5}}}{1}
      {6}{{{\color{numb}6}}}{1}
      {7}{{{\color{numb}7}}}{1}
      {8}{{{\color{numb}8}}}{1}
      {9}{{{\color{numb}9}}}{1}
      {:}{{{\color{punct}{:}}}}{1}
      {,}{{{\color{punct}{,}}}}{1}
      {\{}{{{\color{delim}{\{}}}}{1}
      {\}}{{{\color{delim}{\}}}}}{1}
      {[}{{{\color{delim}{[}}}}{1}
      {]}{{{\color{delim}{]}}}}{1},
}

\def\fstar{$\text{F}^\star$\xspace}

\def\haclstar{$\text{HACL}^\star$\xspace}
\def\evercbor{{\ifanon{VERCOR}\else{EverCBOR}\fi}\xspace}
\def\evercddl{{\ifanon{VERCDL}\else{EverCDDL}\fi}\xspace}

\def\groupor{\mathop{/\mkern-5mu/}}
\def\pulseparse{{\ifanon{SEPARSE}\else{PulseParse}\fi}\xspace}

\begin{document}

\title{Secure Parsing and Serializing with Separation Logic\\Applied to CBOR, CDDL, and COSE}
\renewcommand\shorttitle{Secure Parsing and Serializing with Separation Logic Applied to CBOR, CDDL, and COSE}

\author{Tahina Ramananandro}
\email{taramana@microsoft.com}
\author{Gabriel Ebner}
\email{gabrielebner@microsoft.com}
\author{Guido Mart\'inez}
\email{guimartinez@microsoft.com}
\author{Nikhil Swamy}
\email{nswamy@microsoft.com}
\affiliation{%
  \institution{Microsoft Research}
  \city{Redmond}
  \state{WA}
  \country{USA}
}

\begin{abstract}
Incorrect handling of security-critical data formats, particularly in low-level
languages, are the root cause of many security vulnerabilities. Provably correct
parsing and serialization tools that target languages like C can help. Towards
this end, we present \pulseparse, a library of verified parser and serializer
combinators for non-malleable binary formats. Specifications and proofs in
\pulseparse are in separation logic, offering a more abstract and compositional
interface, with full support for data validation, parsing, and serialization. 
\pulseparse also supports a class of recursive formats---with a focus on
security and handling adversarial inputs, we show how to parse such formats with
only a constant amount of stack space.

We use \pulseparse at scale by providing the first formalization of CBOR, a
recursive, binary data format standard, with growing adoption in various
industrial standards. We prove that the deterministic fragment of CBOR is
non-malleable and provide \evercbor, a verified library in both C and Rust to
validate, parse, and serialize CBOR objects implemented using \pulseparse. Next,
we provide the first formalization of CDDL, a schema definition language
for CBOR. We identify well-formedness conditions on CDDL
definitions that ensure that they yield unambiguous, non-malleable formats, and
implement \evercddl, a tool that checks that a CDDL definition is well-formed,
and then produces verified parsers and serializers for it.

To evaluate our work, we use \evercddl to generate verified parsers and
serializers for various security-critical applications. Notably, we build a
formally verified implementation of COSE signing, a standard for cryptographically
signed objects. We also use our toolchain to generate verified code for other
standards specified in CDDL, including DICE Protection Environment, a secure
boot protocol standard.
We conclude that \pulseparse offers a powerful new foundation on which to build
verified, secure data formatting tools for a range of applications.
\end{abstract}



\maketitle

\section{Introduction}

\begin{figure}
\hspace*{-2em}\includegraphics[trim={0 2.5cm 8cm 0.8cm},clip,scale=.35]{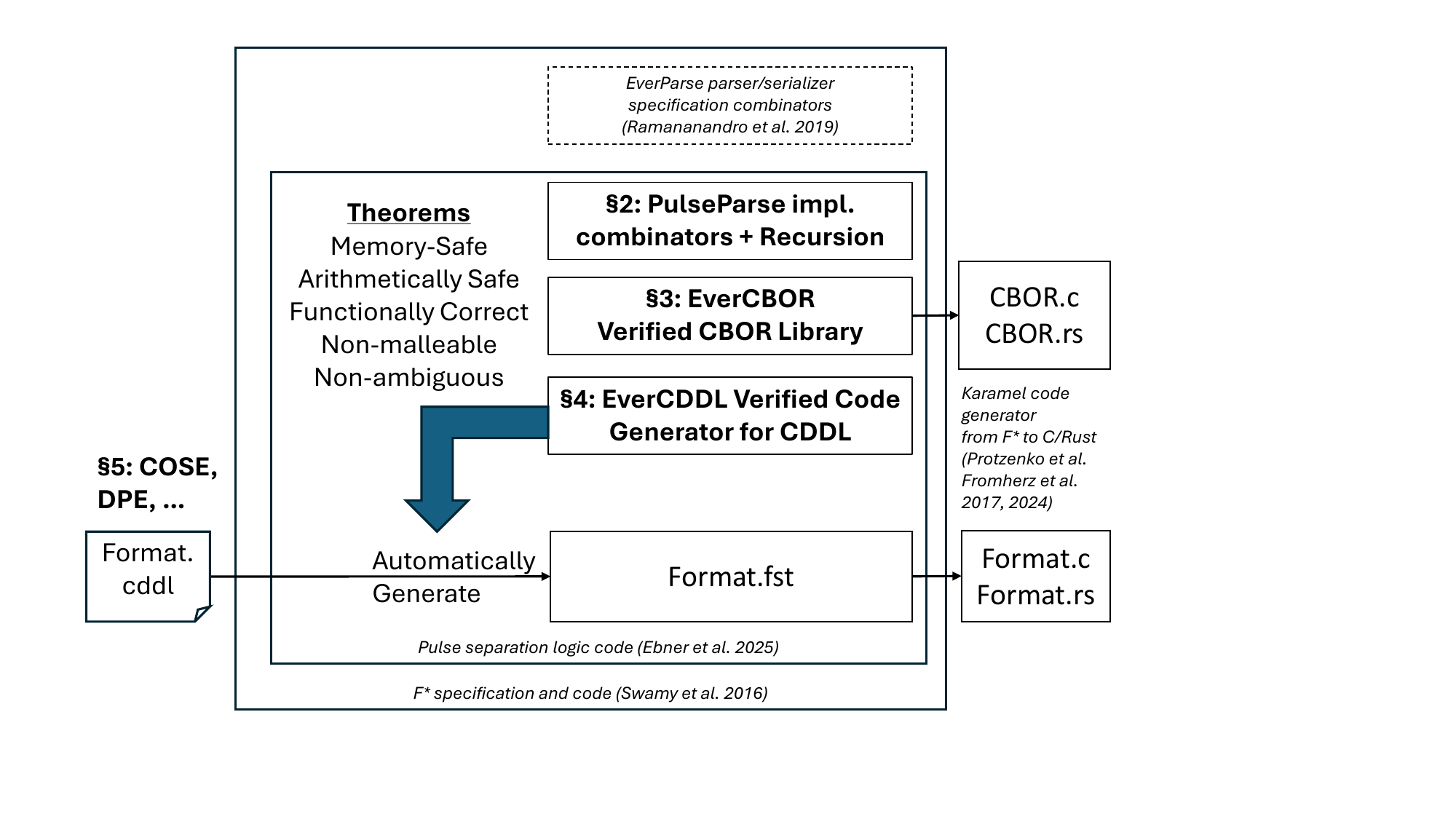}
\caption{Architecture of our contributions}
\label{fig:arch}
\end{figure}

Incorrect handling of security-critical data formats, be it in parsing attacker
controlled data, or in serializing data for cryptographic applications,
is a major source of security
vulnerabilities~\cite{finney2006bleichenbacher,pyrsa}. In response, there is a
rich area of research into tools for secure
parsing~\cite{nailosdi2015,parsley20,daedalus24,everparse,costar21}. We are
particularly interested in secure handling of binary data formats for use in
security-critical low-level applications, in C and other systems programming
languages, including in OS components, embedded systems, and in cryptographic
applications.
 
In this context, formally proven parser generators have been used to secure
critical, commercial software including in Microsoft's OS and cloud
infrastructure~\cite{everparse3d}. However, such uses have focused primarily on
validating flat, tag-length-value encodings of network packet formats. We aim to
broaden the scope of secure, low-level binary formatting tools, enabling them to
handle richer formats (such as those with certain forms of recursion) and to
flexibly support both parsing and serialization, in a performant,
zero-copy-by-default, low-level style.

Our \emph{first contribution} (\S\ref{sec:pulseparse}) is \pulseparse, a new
verified library for secure parsing and serialization. \pulseparse is
implemented in \fstar~\cite{fstar} and in its separation logic sub-language
Pulse~\cite{pulsecore}, with formal proofs of memory safety, functional
correctness, and non-malleability (i.e., unique binary representation) of
formats. The design of \pulseparse employs a novel application of separation
logic to parser \& serializer combinators, yielding an abstract style of
specification with compositional proofs. \pulseparse also supports a class of
security-relevant recursive formats, namely those that can be validated in
constant stack space.

Recursion in \pulseparse is essential to model CBOR (Concise Binary Object
Representation) \cite{cbor}, an Internet standard for the binary representation
of general-purpose JSON-like data structures. A subset, Deterministically
Encoded CBOR, aims to offer non-malleability, thus avoiding hashing-based
authentication bugs that have occurred in similar binary
formats~\cite{decker2014bitcoin}. Our \emph{second contribution}
(\S\ref{sec:cbor}) is \evercbor, a formalization of CBOR, including a proof that
its deterministic encoding is indeed non-malleable---the first such proof.
Implemented in \pulseparse, \evercbor produces verified code in both C and safe Rust
for validating, parsing, and serializing CBOR objects.

CBOR is a single uniform data format, deferring data specifications to a
high-level schema language for CBOR items, called CDDL (Concise Data Definition
Language) \cite{cddl}. By splitting those two concerns, CBOR and CDDL greatly
help protocol designers specify data schemas with extensibility and
forward-compatibility in mind. Our \emph{third contribution} (\S\ref{sec:cddl})
is a formalization of CDDL, including inferring well-formedness conditions on
CDDL definitions (in the form of a new elaboration algorithm, proven sound) that
yield unambiguous and non-malleable formats. Our formalization takes the form of
a tool, \evercddl, that first formally proves that a CDDL definition is
well-formed, and then generates a custom data type along with corresponding
low-level parsers and serializers in \pulseparse, formally verified for inverse,
non-malleability, memory safety, and functional correctness with respect to the
CBOR and CDDL specifications.

CDDL is used in dozens of other standards, in applications including
supply-chain integrity (SCITT)~\cite{scitt}, device attestation protocols
(DPE)~\cite{dpe}, and WebAuthn passwordless authentication \cite{webauthn}. Perhaps its most
prominent use is in the specification of COSE (CBOR Object Signing and
Encryption)~\cite{cose}, a standard for cryptographically signed and encrypted
objects, certificates, and keys, itself used in security-critical
applications such as SCITT, DPE, Client-to-Authenticator Protocol (CTAP) \cite{ctap2-2,fido-proof},
and vaccine certificates~\cite{cddlcovid}.
Our \emph{fourth contribution}
(\S\ref{sec:experiments}) is to evaluate our libraries on two applications, COSE
and DPE. First, we show how to adapt the CDDL specifications of COSE and DPE so
that they are provably unambiguous and non-malleable (using \evercddl), and then
integrate the resulting parsers and serializers in verified applications. For
COSE, we produce a verified library for COSE signing, relying on verified
cryptographic implementations from \haclstar~\cite{haclstar}, proving that the
payload of a signature object is exactly the signature of the to-be-signed bytes
with the given key. For DPE, we show how to integrate our verified parsers and
serializers with a prior verified implementation~\cite{pulsecore}.

Figure~\ref{fig:arch} shows the overall architecture of our contributions. All
the theorems in this paper and the software described are formally verified in
\fstar \cite{fstar}, with the separation logic parts developed in
Pulse~\cite{pulsecore}, an embedded language in \fstar. All our code is proven
memory safe, arithmetically safe, and functionally correct, and the formats we
formalize are all proven non-malleable. Our software can be used in verified
Pulse applications, as we do for COSE and DPE. Additionally, using Karamel, an existing
code generator~\cite{lowstar,krml2rust}, \evercbor extracts from Pulse to a
standalone library in C and in safe Rust, with idiomatic, defensive APIs as a drop-in
replacement of existing unverified CBOR libraries used in a variety of
applications. We evaluate \evercbor against commonly used unverified CBOR libraries such as QCBOR~\cite{qcbor} and TinyCBOR~\cite{tinycbor}, noting that we support more CBOR features (including
arbitrary maps), and implement all necessary checks, find that our verified code
is competitive in speed and memory consumption. Verified code produced by
\evercddl also extracts to a standalone library in either C or safe Rust.

All the artifacts we contribute \ifanon{have been renamed for double-blind
submission}\else{have been merged into EverParse\footnote{\label{footnote:everparse}\url{https://github.com/project-everest/everparse/}}}\fi.
Note, throughout the paper, we say "format" to mean "parsing and serialization",
e.g., we say "format combinators" to mean "parser and serializer combinators".

\section{\pulseparse: Format Combinators with Separation Logic}
\label{sec:pulseparse}

Combinator parsing has its roots in functional programming~\cite{Hutton89},
providing a higher-order, compositional way to structure parsers. We seek to use
combinator parsing to produce verified code in low-level languages, a technique
leveraged first by EverParse~\cite{everparse}, a format combinator library in
\fstar with formal proofs of correctness and security, yielding verified C code.
Their verification approach is to layer the combinators, distinguishing between
\emph{specification} combinators, pure functions that \emph{define} the data
format specification, and on which proofs of properties such as non-malleability
are conducted; and \emph{implementation} combinators which follow the structure
of the specification combinators while refining them to efficient, low-level
code.

\pulseparse follows this approach too. In fact, we simply reuse many of the
specification combinators from EverParse, though we contribute some new
specification combinators, notably for recursive formats. Our first main
innovation is a new library of implementation combinators, whose proofs are
structured using separation logic, contrary to EverParse, which uses a classical
Hoare logic. As we will explain, through the use of separation logic,
\pulseparse proofs are more modular, and enable abstract implementation
combinators, simplifying both their construction and, more importantly, proofs
of their clients.

\subsection{Specification Combinators (Review)}

A \emph{parser specification} in \pulseparse is a pure \fstar function of type \
\lsf{parser t = seq U8.t -> option (t & nat)} (with an extra non-malleability condition as below),
that takes as argument a sequence
of input bytes, and returns \lsf{Some(v, n)} if parsing succeeds and the first
\lsf{n} input bytes are a binary representation of the high-level value \lsf{v};
and \lsf{None} if parsing fails.

Such a parser specification defines the data format, and properties about the
format can be proven as lemmas. EverParse parser specifications are
required to be \emph{non-malleable}: for a given data format defined by its parser
specification \lsf{p}, if \lsf{p(b1) = Some (v, n1)} and 
\lsf{p(b2) = Some (v, n2)}, parsing two input byte sequences \lsf{b1} and
\lsf{b2} to the same high-level value \lsf{v}, then \lsf{n1 = n2} and \lsf{b1}
and \lsf{b2} coincide on their first \lsf{n1} bytes, meaning that the first
\lsf{n1} bytes of \lsf{b1} are a unique binary representation for \lsf{v}.
Non-malleability is especially important for security-critical applications,
especially in cryptographic contexts---several prominent attacks come down to
malleability of formats~\cite{finney2006bleichenbacher,pyrsa}.

To combine parsers, e.g., \lsf{p1: parser t1} and \lsf{p2: parser t2}, one uses
a combinator \lsf{parse_pair p1 p2: parser (t1 & t2)}, a parser specification for
a pair of values whose binary representations are laid out side-by-side. Since
\ls`p1` and \ls`p2` are non-malleable, one can prove that \lsf{parse_pair p1 p2}
is also non-malleable by construction.

Given a parser specification \lsf{p: parser t}, a serializer specification for
\lsf{p} has type \lsf{serializer p = (x: t) -> (b: seq U8.t \{ p b == Some (v,
length b) \})}: serializing a high-level value \lsf{(v)} yields a sequence of
bytes guaranteed to parse back to \lsf{v}, specified with a \emph{refinement
type} on the return value. Serializers can also be combined, e.g., given
\lsf{s1: serializer p1} and \lsf{s2: serializer p2}, the pair serializer
\lsf{serialize_pair s1 s2}
serializes a \lsf{(v1, v2): t1 & t2}, by running \lsf{s1} on \lsf{v1}, then
\lsf{s2} on \lsf{v2}, and concatenate the resulting byte sequences. The
combinator \lsf{serialize_pair s1 s2} has type \lsf{serializer (parse_pair p1
p2)} only if \lsf{p1} has the \emph{prefix property}: for any sequence of input
bytes \lsf{b} such that \lsf{p(b)} returns \lsf{Some(v, n)}, \lsf{p(b')} returns
the same result for any input \lsf{b'} coinciding with \lsf{b} on its first
\lsf{n} bytes. This property is necessary to prove the correctness of
\lsf{serialize_pair} with respect to \lsf{parse_pair}.

\pulseparse reused EverParse's specification combinators for various types, such
as machine integers, bit fields, value-dependent pairs, lists of a given number
of elements, checking for a value property, and rewriting under a bijection.

\subsection{Implementation Combinators with Separation Logic}

For implementation combinators, \pulseparse uses \emph{separation logic}
\cite{seplogic}, as provided by Pulse \cite{pulsecore}. For a given pair of
parser and serializer specification combinators, we implement several
combinators in Pulse: \emph{validators}, \emph{jumpers}, \emph{accessors} and
\emph{readers} for parsing; and \emph{writers} for serialization. 

\paragraph{Validators} For \lsf{p:parser t} and \lsf{s:serializer p},  a
\emph{validator} \lsf{v} is a Pulse function (i.e., a procedure with possible
side effects) that takes an input byte array and returns the number of bytes
consumed by \lsf{p} if parsing succeeds, or an error code if parsing fails.
Whereas \lsf{p} is specified on an input sequence of bytes, \lsf{v} reads the
contents of a concrete array stored in memory, recording the number of bytes
read in a mutable out parameter storing a machine integer \lsf{U64.t}, and
returning true if, and only if, \lsf{p} would have succeeded on the abstract
byte contents of the array. Validators can also be combined (e.g.,
\lsf{validate_pair v1 v2}). \fstar inlines the combinator definition when
transpiling to C or Rust, so that the resulting code is first-order.

\paragraph{Jumpers} A \emph{jumper} is a Pulse function that takes an input byte
array required to start with a valid byte representation with respect to
\lsf{p}, and returns the number of bytes consumed---it is used to "jump" over a
known valid item in a byte array.

\paragraph{Accessors} An \emph{accessor} is a Pulse function that takes an input
byte array containing a valid byte representation, and returns a (pointer to a) subarray
containing the valid byte representation of a subobject.
\pulseparse is careful to
ensure that no heap accesses are incurred in the process. Accessors specified in
separation logic can be pleasingly abstract. Consider for instance the
\lsf{parse_pair} combinator for parsing a pair of data. For \lsf{p1:parser t1},
\lsf{p2:parser t2}, \lsf{s1:serializer p1} and \lsf{s2:serializer p2},  to
implement a pair accessor, we need the jumper \lsf{j1} for \lsf{p1}, to jump
over the first component of the pair.
Then, for a byte array \lsf{a} and a pair \lsf{(v1:t1, v2:t2)}, we introduce a
separation logic predicate, \lsf{ser (serialize_pair s1 s2) a (v1, v2)},
stating that the current contents of the byte array \lsf{a} is exactly the byte
representation of \lsf{(v1, v2)} obtained by the corresponding pair serializer
specification. Then, we specify a call to a pair accessor implementation in
Pulse using the following separation logic triple---this is our first glimpse of
separation logic and we explain the specification in detail below.
\begin{lstlisting}[language=fstar]
{ ser (serialize_pair s1 s2) a (v1, v2) }
  let (a1, a2) = access_pair j1 a
{ (ser s1 a1 v1 * ser s2 a2 v2) *
 ((ser s1 a1 v1 * ser s2 a2 v2) -* ser (serialize_pair s1 s2) a (v1, v2)) }
\end{lstlisting}

The first part of the triple is the \emph{precondition}, which describes the
relevant part of memory as a separation logic proposition (of type \lsf{slprop})
required to hold before running the Pulse statement mentioned in the middle part
of the triple. In this case, the precondition simply states that before running
the accessor, one must prove that the array \lsf{a} contains a valid
serialization of \lsf{(v1, v2)}.

The last part of the triple is the \emph{postcondition}, describing a property
of the memory upon completion of the Pulse statement. The postcondition uses two
separation logic connectives. First, \lsf{A * B} is a \emph{separating
conjunction}, meaning \lsf{A} and \lsf{B} are separation logic predicates that
describing disjoint ownership of memory. Owning the predicate \lsf{A} means no
other part of the program can disturb the validity of \lsf{A}. The postcondition
also uses a \emph{magic wand}\footnote{Our proofs use Pulse's \emph{trades}, which have a slightly different semantic model than magic wands, but we only rely on usual proof rules that are valid for both the traditional magic wand and Pulse trades.}, \lsf{A -* B}, which is a connective that enjoys
an \emph{elimination proof rule} 
$\frac{\mbox{\lsf{A * (A -* B)}}}{\mbox{\lsf{B}}}$. That is, one can trade an
\lsf{A} and (separately) \lsf{A -* B} for a \lsf{B}.

Specifically, in the context of the triple above, the postcondition says that
(1) \lsf{a1} points to an array segment that contains a valid serialization of
\lsf{v1}; (2) \lsf{a2} points to an array segment that contains a valid
serialization of \lsf{v2}; and (3) one can give up ownership of \lsf{a1} and
\lsf{a2} to recover ownership of the  entire original array segment \lsf{a} that
contains a serialization of \lsf{(v1, v2)}. This specification captures the
essence of \emph{zero-copy} parsing with no heap allocation---the caller gains
ownership to the relevant array segments, and when it is done with them, it can
simply relinquish ownership and use the magic wand to regain ownership to the
original array \ls{a}.

Note how this specification makes no mention of array offsets, which are
abstracted away. Using a similar pattern, we provide an accessor for dependent
pairs, accessors for the head and the tail of a nonempty list of elements. The
pattern \lsf{A * (A -* B)} is common enough that we abbreviate it as \lsf{A >*
B}.

\paragraph{Readers}
For base values such as integers, we provide \pulseparse \emph{readers}, to
return the actual values. Given a parser-serializer specification pair \lsf{p:
parser t}, \lsf{s: serializer p}, a reader applied to an array \lsf{a} satisfies
the following triple:
\begin{lstlisting}[language=fstar]
  { ser s a v } let v' = reader s a { ser s a v * (v' == v) }
\end{lstlisting}
showing that it returns a value \ls{v'} equal to the value \ls{v} from the
precondition, without changing the array \lsf{a}. 

Now, for \lsf{r1: reader s1}, \lsf{j:jumper p1}, and \lsf{r2: reader s2}, we
can implement \ls{read_pair r1 j1 r2 : reader (serialize_pair s1 s2)}:
\begin{lstlisting}[language=fstar]
  let (x1, x2) = access_pair j1 x;
  let v1 = r1 x1; let v2 = r2 x2;
  wand_elim _ _; (* a ghost proof step for the elimination rule for -* *)
  (v1, v2)
\end{lstlisting}
Unlike in EverParse, \pulseparse requires no offset reasoning.

In \pulseparse, we also provide value readers for dependent pairs and
bitfields. However, we do not provide value readers for lists or other
variable-sized data, since we want to parse data without heap
allocating, and only using constant stack space with respect to the
input, which may be attacker-controlled.

Rather, we provide \emph{zero-copy readers}, which are functions that ``save''
the pointers to subcomponents without parsing them. To this end, for a \lsf{p:
parser t}, and \lsf{s: serializer p}, one needs to choose a low-level datatype
representation \lsf{u} into which to parse such data, along with a separation
logic predicate \lsf{r: u -> t -> slprop}. Then, a zero-copy reader is a Pulse
function of the following signature:
\begin{lstlisting}[language=fstar]
{  ser s a v } let res = zerocopy s r a { r res v >* ser s a v }
\end{lstlisting}
Similarly to accessors, the postcondition uses the \emph{magic wand}, thus
allowing to ``borrow'' permissions from subpointers of \lsf{x} into \lsf{res},
controlled by \lsf{r}. For instance, a zero-copy reader can read a pair of a
machine integer and a sequence of bytes, by returning the value of the integer
and a pointer to the byte array, which it will thus not copy. Then, the
application can use further accessors and readers to manipulate that byte array.
We provide several zero-copy reader combinators, e.g., a value reader and a
value-dependent zero-copy reader can be combined together to obtain a zero-copy
reader for a dependent pair of values.
\iffull
In \pulseparse, we provide the following zero-copy reader combinators:
\begin{itemize}
\item A value reader can be lifted as a zero-copy reader, by setting the low-level type \lsf{u} to be identical to the high-level type \lsf{t}, and \lsf{r x y = pure (x == y)}.
\item A byte array valid with respect to some serializer specification \lsf{s} can be left as is: \lsf{u = byte\_array} and \lsf{r x y = ser s x y}.
\item Two zero-copy readers can be paired together to obtain a zero-copy reader for a pair of values.
\item A value reader and a value-dependent zero-copy reader can be combined together to obtain a zero-copy reader for a dependent pair of values.
\item The high-level values read by a zero-copy reader can be rewritten with a bijection without changing the low-level value returned by the zero-copy reader.
\item A zero-copy reader for low-level type \lsf{u1} and relation \lsf{r1} can be turned into a zero-copy reader for low-level type \lsf{u2} and relation \lsf{r2} and the same high-level values, if provided a Pulse function going from \lsf{u1} to \lsf{u2}:
\begin{lstlisting}[language=fstar]
(x1: u1) -> (y: t) -> stt u2 (requires r1 x1 y)
(ensures fun (x2: u2) -> r2 x2 y ** (r2 x2 y -* r1 x1 y))
\end{lstlisting}
\end{itemize}
\fi
We provide no zero-copy reader that would allocate into the heap. In
particular, we do not allocate references, byte arrays, or other kinds
of arrays.

\paragraph{Writers}
Similarly to value readers, we provide value writer combinators,
taking a value to write, and an output byte array, and returning the
number of bytes written. We also provide value size combinators to
compute the minimal size required for serialization; these value size
combinators take a bound as argument, and gracefully fail if the size
needed is larger than the bound, so as to avoid arithmetic overflows.

However, contrary to input, an application may want to build data structures for
which it entirely controls nesting and memory usage, apart from byte arrays
containing unparsed data. For this, we define \emph{copy writers} for generic data
structures so that an application can define such a data structure, populate it
in any order, and serialize it all at once by copying it into an output buffer
following a serializer specification. Then, we specify copy writers as:
\begin{lstlisting}[language=fstar]
{ exists w. out |-> w * r vl vh * (length w <= length (s vh)) }
let res : size_t = writer s r a out
{ exists w'. out |-> (append (s vh) w') * r vl vh }
\end{lstlisting}
where \lsf{s} is a serializer specification, \lsf{r : tl -> th -> slprop} is a
relation between the actual data structure \lsf{vl:tl} that the application
built and wants to serialize and \lsf{vh:th}, some abstract specification-level
value. A writer expects as its precondition that the application has proven
\lsf{vl} and \lsf{vh} are related, and that the output buffer is large enough to
contain the serialized bytes \lsf{s(y)}. 

The separation logic predicate \lsf{r} plays a critical role in the correctness
of the \pulseparse copy writers. For instance, \lsf{tl} can be the Pulse type
\lsf{ref th} of non-null pointers to values of type \lsf{th}, in which case
\lsf{r vl vh} will be \lsf{vl $\,\mapsto$ vh}, the predicate saying that the
contents of the reference \lsf{vl} is \lsf{vh}. Then, that writer will read the
contents of \lsf{vl}, which is equal to \lsf{vh}, and then call a value writer.
We provide many copy writer combinators, including, for example, given two copy
writers for low-level types \lsf{tl1} and \lsf{tl2} and two separation logic
predicates for the same high-level type \lsf{th}, we provide a copy writer for
low-level type \lsf{tl1 + tl2}, thus allowing several low-level data structures
for the same high-level type. Lacking support for abstract relations between low
and high-level representations, serialization in EverParse requires applications
to directly write into the output buffer in the right order, incurring heavy
application-level reasoning about output offsets.

\iffull
More generally, in \pulseparse, we define the following copy writer
combinators:
\begin{itemize}
\item we lift value writers to copy writers using \lsf{th = tl} and \newline \lsf{r x y = pure (x == y)}
\item we provide a byte-copy writer for byte arrays containing unparsed bytes valid with respect to serializer specification \lsf{s} (with \lsf{tl = byte_array} and \lsf{r x y = ser x y})
\item given a copy writer for low-level type \lsf{tl}, we provide a copy writer for low-level type \lsf{ref tl}
\item given a copy writer for low-level type \lsf{tl}, we provide a copy writer for low-level type \lsf{array tl n}  to serialize an array of $n$ elements.
\item given two copy writers for low-level types \lsf{tl1} and \lsf{tl2}, we provide a pair copy writer for low-level type \lsf{tl1 & tl2}
\item given two copy writers for low-level types \lsf{tl1} and \lsf{tl2} and two separation logic predicates for the same high-level type \lsf{th}, we provide a pair copy writer for low-level type \lsf{tl1 + tl2}, thus allowing several low-level representations for the same high-level type.
\item given a value writer for the left-hand-side of a value-dependent pair, and a copy writer for the right-hand-side, we provide a value-dependent pair copy writer
\end{itemize}
\fi

In Section~\ref{sec:cddl}, we define a set of CDDL serializer combinators using
a similar methodology, which we extend to define parser combinators using
low-level representations that can contain both application-controlled data
structures and user-controlled unparsed data. Based on our learnings from CDDL, 
we integrated similar parsing combinators in \pulseparse---Appendix~\ref{appendix:arith}
shows them in use on a small recursive format for arithmetic expressions.

\paragraph{Support for some recursion}

Most memory-constrained binary data parsers do not support arbitrary
recursion, because many recursive formats require a stack or other
form of memory whose size would grow with the input, thus exposing
themselves to attackers exhausting memory during validation or
parsing.

However, there is a class of recursive data formats that allow validation in
constant stack and memory space---we will see in Section~\ref{sec:cbor} that
CBOR belongs to this class.

\begin{theorem}\label{thm:recursive-formats-constant-stack}
Consider a binary data format where an object representation starts
with a header, followed by a contiguous sequence of recursive object payload
entries, and nothing afterwards. If a header can be validated in
constant stack and memory space, and if the header of an object alone
is enough to determine the number of immediate children of this
object, then data in this format can be validated in constant stack
and memory space.
\end{theorem}

To support this class, we include in \pulseparse a recursive parser
specification combinator \lsf{parse_rec} as follows: let \lsf{th} be the
high-level type of headers, \lsf{t} be the high-level type of objects, \lsf{p} a
parser specification for headers consuming at least one byte, \lsf{count} a
function computing the number of recursive payload elements,
and \lsf{synth} a function synthesizing a high-level object from its header and elements;
then, \lsf{parse_rec} is defined below, where
\lsf{let!} sequences \lsf{option} computations:
\begin{lstlisting}[language=fstar]
let rec parse_rec' (ph:parser th) (count:th -> nat) 
  (synth: (h: th) -> (l: nlist t (count h)) -> t) (fuel: nat) (b: Seq.seq U8.t)
: option (t & nat)
= if fuel = 0 then None else
   let! h, size = ph b in (* parse th header *)
   let b' = slice_from b size in (* b' contains (count h) elements to be parsed *)
   let! l = parse_nlist (parse_rec' ph count synth (fuel-1)) (count h) b' in
   Some (synth h l) (* map the parsed values to a high-level value t *)
let parse_rec ph count synth b = parse_rec' ph count synth (1+length b) b
\end{lstlisting}

We prove that if a \lsf{ph} header has the prefix property, then so does
\lsf{parse_rec ph count synth},
and if \lsf{synth} is injective and \lsf{ph} is non-malleable, then
\lsf{parse_rec ph count synth} is non-malleable.

Of course, \lsf{parse_rec} is recursive---but it is only a specification
combinator and not meant to be executed. The implementation combinators use only
constant stack.\footnote{ Note, stack space usage is outside the scope of our
formal proof, since the underlying logic does not provide a way to specify it.
However, we only use while loops and use no recursive functions, so the function
call depth is bounded statically (by the number of function definitions.)  
} For validation, we take as argument a header validator, and a function to
retrieve the number of expected payload items in the payload. Then we maintain a
counter of expected items, which we initialize to 1. Whenever we start
validating an item, we decrease that counter. Then we add the number of expected
items in the payload. The validator succeeds if the counter reaches 0. Using
Pulse, we prove that our validator is functionally correct with respect to its
recursive specification, using a loop invariant. We take care of avoiding
arithmetic overflow by leveraging the fact that a valid header always consumes
at least one byte. See Appendix~\ref{appendix:cbor-validator} for more details;
Appendix~\ref{appendix:arith} provides an example \pulseparse for a recursive
format of variable-arity trees.

We now turn to our formalization of CBOR, making essential use of \pulseparse's
support for recursion, and abstract separation logic specifications.

\section{\evercbor: A Verified Generic CBOR Parser and Serializer} 
\label{sec:cbor}

JSON is a ubiquitous textual representation of data. However, it comes with a
vast collection of issues, some related to efficiency (whitespace, decimal
integers, etc.), others related to security (parsing errors due to bad nesting
of quotes or braces, string injection, etc.)  This is why Internet practitioners
have long sought binary representation alternatives, such as UBJSON, BSON, or
MessagePack.\footnote{\url{https://ubjson.org/}, \url{https://bsonspec.org/},
\url{https://msgpack.org/}} For uniformity and extensibility reasons, the IETF
adopted CBOR as an Internet Standard in 2020 as RFC~ 8949 \cite{cbor}. Since
then, CBOR rapidly evolved into a binary format of its own, defining its own set
of items extending JSON.

\subsection{Background: CBOR} \label{sec:cbor:data-model}

A CBOR item \iffull (Figure~\ref{fig:cbor-data-model}) \fi can be any one of: a
64-bit nonnegative integer, a 64-bit negative
integer represented as ``one's complement'',
a ``simple value'', a byte string, a UTF-8 text string, a CBOR item tagged with
a nonnegative 64-bit integer, a finite array of CBOR items (a heterogeneous
ordered sequence,) or a finite key-value map, where each entry \emph{key} or
value can be an arbitrary CBOR item---a generalization of JSON, where only
strings are allowed as keys. \iffull Moreover, simple values are a subset of
non-negative byte values, meant to generalize JSON's Boolean type by encoding
symbols such as NULL, meant to be distinct from integer values or empty strings.
\fi

\iffull
\begin{figure}
\[
\begin{array}{rcl}
\textrm{cbor} & ::= & \textsf{Int} ~ (x \in [-2^{64}; 2^{64}-1]) \\
& | & \textsf{Simple} ~ (x \in [0, 23] \cup [32, 255]) \\
& | & \textsf{ByteString} ~ (n \in [0, 2^{64} - 1], x \in [0, 255]^n) \\
& | & \textsf{TextString} ~ (n \in [0, 2^{64} - 1], x \in [0, 255]^n \cap \textrm{UTF-8}) \\
& | & \textsf{Tagged} ~ (\textrm{tag} \in [0, 2^{64} - 1], v: \textrm{cbor}) \\
& | & \textsf{Array} ~ (n \in [0, 2^{64} - 1], x \in \textrm{cbor}^{n}) \\
& | & \textsf{Map} ~ (n \in [0, 2^{64} - 1], x: ~ (\textrm{cbor} \xrightarrow{n} \textrm{cbor}))
\end{array}
\]
where $(t \xrightarrow{n} t)$ is the type of extensional maps with $n$ entries
\caption{The data model for CBOR items}
\label{fig:cbor-data-model}
\end{figure}
\fi

However, na\"ively transcribing the above description as a grammar or an
inductive datatype could conflate maps with lists of pairs of items, potentially
allowing duplicates in map keys. Thus, secure applications must make sure CBOR
maps have no such duplicates, to avoid misunderstandings where different
applications will look at different entries for one given key. Moreover, unlike
arrays, map entries are unordered. Thus, the entry keys of a map are better
modeled as a set rather than as a list. These problems are inherited from JSON.
Further, trying to directly define CBOR items as an inductive type in a proof
system is not possible, since such a definition would require the type to appear
negatively in map keys. So, we look to the byte representation as a basis for
formalizing CBOR's data model.

\begin{figure}
\hspace*{-2em}\includegraphics[trim={0 1cm 0 1cm},scale=.35]{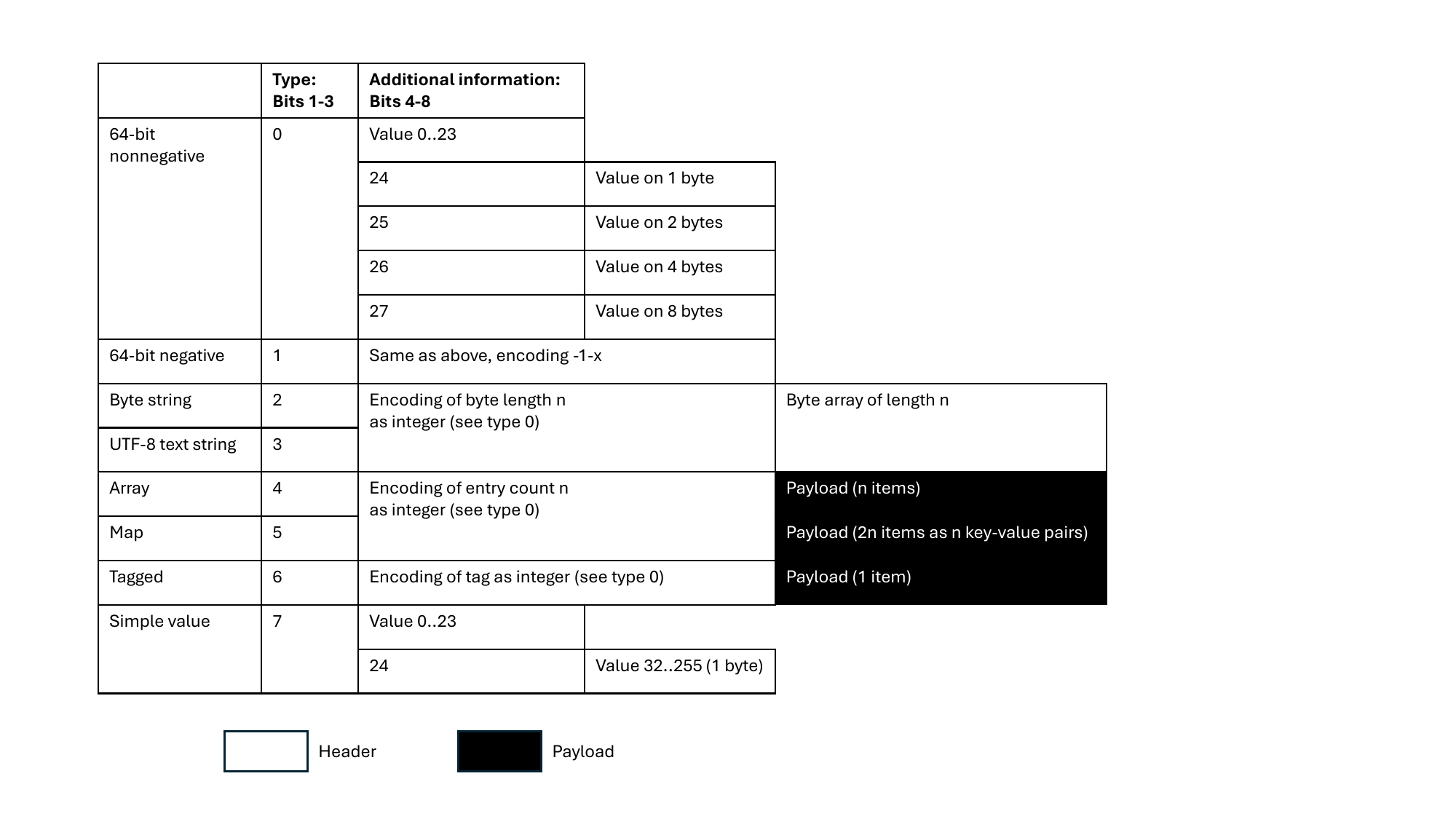}
\caption{Representing CBOR items as bytes}
\label{fig:cbor-binary}
\end{figure}

CBOR defines a byte representation for its items in a tag-count-payload
fashion. Figure~\ref{fig:cbor-binary} shows how CBOR items are
represented as bytes. The first byte contains three bit fields, the
most significant 3 bits of which describe the type of the CBOR item. The
remaining 5 bits, called ``additional information'', encode an integer from 0
to 31: additional info 0 to 23 encode
a nonnegative integer (or a simple value) of this value. For 64-bit
integers, info 24, 25, 26 and 27 encode the fact that the integer is
encoded in the next 1, 2, 4 and 8 bytes respectively. Thus, integers
are encoded in variable length. Byte and text strings are prefixed
with their byte size as a 64-bit integer encoded in the same way
(starting from the 4th bit of the first byte), thus limiting their
byte size to $2^{64} - 1$. Similarly, arrays and maps start with
their number of entries as a 64-bit integer
(thus limiting their entry count to $2^{64} - 1$,) followed by their
entries consecutively, where a map entry consists in two consecutive
CBOR items; and tagged items start with their tag as a 64-bit integer,
followed by the payload item. Type number 7 is used for simple
values.

Not all such binary data represent valid CBOR items. Once binary data conforms
with this representation, a validator needs to check for the absence of map key
duplicates. We call \emph{raw CBOR bytes} any sequence of bytes conforming to
the binary representation but not yet checked for the absence of map key
duplicates.

\paragraph{Deterministically-Encoded CBOR}
A given CBOR item has several possible representations, owing to the variable
byte size of integer values and length prefixes, and the order in which map
entries are serialized. This allows for \emph{malleability attacks:} if an
application cryptographically signs the byte representation of a CBOR item, an
attacker
could possibly construct a \emph{different} representation
of the same item, which the application would not recognize as having signed.
This can lead to serious security issues, see e.g.~\cite{decker2014bitcoin}.
To prevent this issue, Deterministically Encoded CBOR \cite[\S
4.2.1]{cbor}, mandates integers to be serialized in their shortest form, and map
entries to be serialized in the increasing lexicographic order of the byte
representations of their keys. We prove that this subset indeed provides a
unique binary representation for all CBOR items. Deterministically Encoded CBOR
is used in COSE and many other security-critical protocols requiring unique
binary representation.

\iffull
\paragraph{Indefinite-length CBOR byte representations}
CBOR can also represent arrays and maps in an indefinite-length way,
where, instead of storing the number of entries in the prefix of their
byte representation, indefinitely many entries can be parsed until a
special ``stop'' item is encountered. Such representations are
explicitly excluded from the deterministic subset of CBOR, which
requires all arrays and maps to have definite lengths specified in
their byte representation prefixes.

Even beyond the deterministic subset, we believe that
indefinite-length representations introduce several security
issues. Such representations are mostly meant for processing of
streamed input data. Implementations could start processing such
streamed data before having reached its end, and thus, an attacker
could induce unexpected behavior by providing a CBOR data stream that
starts with valid input and suddenly becomes invalid. This is why we
choose to validate all bytes at once after making sure we receive it
all, and start processing them only once we make sure the whole
representation is valid.

With our choice, a CBOR byte representation validator accepting
indefinite-length maps and arrays arbitrarily nested with
definite-length maps and arrays needs to store the number of CBOR
items left to be validated after each ``stop'' item. For a given
$n$, validating any $n$-depth nesting of a definite-length array of
size at least 2 containing, as its first element, an indefinite-length
array containing the remainder of the nesting, will incur storing a
stack of size $\Omega(n)$. Thus, if a CBOR validator accepts such
arbitrary nestings, then an attacker might exhaust its memory even
with a \emph{valid} byte sequence. For this reason, we choose not to
support indefinite-length representations at all, and from now on, we only
consider definite-length CBOR byte representations.
\fi

\subsection{Formalizing Raw CBOR in \pulseparse} \label{sec:evercbor}

We start by specifying and implementing a formally verified data validator,
parser and serializer for raw CBOR bytes. Then, we use raw CBOR bytes
specification to describe the CBOR data model, and we reuse the implementations
for the deterministic subset of CBOR.

\paragraph{Specification and input validation}
\label{sec:cbor:raw}

We start with a specification for validating raw CBOR bytes. This can be done in
constant stack space, using our validator for recursive formats, since CBOR
meets the requirements of Theorem~\ref{thm:recursive-formats-constant-stack}.
\iffull
CBOR item always starts with a \emph{header} followed
by a payload of other CBOR items, with nothing in between these CBOR
items; and the header alone is enough to know how many CBOR items need
to be validated in the payload ($n$ for an array of $n$ entries, $2n$
for a map of $n$ entries, $1$ for a tagged item, and $0$ otherwise.)
In fact, the header consists in the first byte, and the additional
bytes needed to encode the number of array or map entries and the tag
of a tagged item. For other items (integers, simple values, strings
), the whole CBOR item is its own header, and the payload is
empty.
\fi
Our formalization starts by first defining a type of \emph{raw CBOR data} shown
(partially) below, representing maps as lists of pairs, and raw integers paired
with a bound on their size in bytes.
\begin{lstlisting}[language=fstar]
type raw_u64 = { size:nat{ size <= 4 }; v:U64.t { fits_in size v }; }
type raw_data =
| Int64: (t:U8.t {t=0 \/ t=1}) -> (v:raw_u64) -> raw_data
 $\ldots$
| Map: (len:raw_u64) -> (v:nlist (raw_data & raw_data) len.v) -> raw_data
\end{lstlisting}
The we apply \lsf{parse_rec} to raw CBOR bytes, where \lsf{count_payload} reads
the header to compute the number of items of each case; \lsf{parse_header} is a
simple parser for the header bytes, and \lsf{synth_payload} constructs a
\lsf{raw_data} from the list of parsed items.
\begin{lstlisting}[language=fstar]
let parse_raw : parser raw_data =
  parse_rec parse_header count_payload synth_cbor $\mbox{\textit{where}}$
let count_payload = $\function$ | Map len _ -> 2 $\times$ len.v | $\ldots$
\end{lstlisting}

\iffull
\begin{figure}
\begin{lstlisting}[language=fstar]
let raw_u64_prop (size:nat) (value:U64.t) =
  if size = 0 then value <= 23
  else value < pow2 (8 $\times$ pow2 (size - 1))
type raw_u64 =
{   size:  nat   { size <= 4 };
    value: U64.t { raw_u64_prop size value };  }
type raw_data =
| Simple: (v:U8.t { v<=23 \/ v>=32 }) -> raw_data
| Int64: (t:U8.t {t=0 \/ t=1}) -> (v:raw_u64) -> raw_data
| String: (t: U8.t {t=2 \/ t=3}) -> (len:raw_u64) ->
    (v: Seq.lseq U8.t len.value
      {t=3 ==> UTF8.correct v}) -> raw_data
| Array: (len:raw_u64) ->
    (v:nlist raw_data len.value) -> raw_data
| Map: (len:raw_u64) ->
    (v:nlist (raw_data & raw_data) len.value) -> raw_data
| Tagged: (tag:raw_u64) -> (v:raw_data) -> raw_data

let parse_header = ...
let count_payload (x: raw_data) = match x with
| Array len _ -> len.value
| Map len _ -> 2 $\times$ len.value
| Tagged _ _ -> 1
| _ -> 0
let synth_cbor = ...
let parse_raw_cbor = parse_rec parse_header count_payload synth_cbor
\end{lstlisting}
\caption{\fstar inductive type for raw CBOR data} \label{fig:cbor-raw-data}
\end{figure}
\fi

Thus, we prove that the parser specification for raw CBOR data is injective: raw
CBOR bytes are a unique representation of raw CBOR data. This is true because
the raw CBOR data type records all integer byte sizes and retains the order of
all map entries. For any \lsf{parse_rec}, \pulseparse by construction provides a
corresponding low-level implementation combinator for validation, and since the
CBOR header validator and the expected payload count function run in constant
stack space, then so does the raw CBOR byte validator.

\paragraph{Parsing}
Concretely, we do not want to parse raw CBOR bytes into \lsf{raw_data}, since
the latter is recursive and doing so would incur heap allocations.
Instead, we provide an implementation-level parser \lsf{iparse_raw} which parses
an input array of bytes into a low-level data structure of type
\lsf{iraw_data}, which contains a partial parse of the input, with all the
recursive occurrences represented simply by pointers into the input array.
As such, we implement verified, incremental, mostly zero-copy parsing, in the
sense that we do not copy variable-size data, but we only copy a constant amount
of memory for one given call of the raw parser: such a call is not recursive and
will only stack-allocate a constant amount of memory.
We provide \emph{accessors} to inspect the contents of an \lsf{iraw_data}, e.g.,
for an array or a map, \lsf{iparse_raw} reads only its entry count, and we
provide an accessor to iterate over the contents: calling the accessor will run
\lsf{iparse_raw} once on the current array entry, or once on the current map
entry key and once on the value.
\iffull
For an integer, \lsf{iparse_raw} reads it.
For a string, \lsf{iparse_raw} reads its length, and provides a pointer to its
payload.
For a tagged item, \lsf{iparse_raw}r reads its tag, and we provide an accessor
to access its payload: calling the accessor will run \lsf{iparse_raw} once on
the payload. 
\fi

\newcommand\relrawcbor{\ensuremath{\uparrow}}

To specify the correctness of \lsf{iparse_raw}, we define a relation
\lsf{(l:iraw_data) $\,\relrawcbor$ (h:raw_data) : slprop}, relating a
low-level partial parse \lsf{l:iraw_data} to a fully parsed high-level value
\lsf{h:raw_data}. The triple below specifies \lsf{iparse_raw}:
\begin{lstlisting}[language=fstar]
{input |-> b * (|b| = n /\ valid(b)) }
let res : iraw_data = iparse_raw (input, n)
{exists (h:raw_data). (res $\relrawcbor$ h >* input |-> b) * parse_raw(b)==h }   
\end{lstlisting}
The precondition says that, before running the parser, \lsf{input} points to
some byte sequence \lsf{b}, and that \lsf{b} is of length \lsf{n} and starts
with valid raw CBOR bytes. The postcondition shows that one gains access to a
low-level result \lsf{res} corresponding to the high-level parse of \lsf{b}, and
can give up access to \lsf{res} to regain ownership of \lsf{input}.

In full generality, our relation \lsf{l $\,\relrawcbor$ r} is equipped with
fractional permissions~\cite{boyland03frac}, allowing shared readable access to
parsed data. So, one can split 
\lsf{res $\,\relrawcbor$ h}, apply an accessor to \lsf{res} by using one
fraction, leaving the other fraction available to apply other accessors if
needed, and reconstitute the original full permission when one no longer needs
the accessed data.

\paragraph{Serialization}
Whereas using accessors on \lsf{iraw_data} is enough to read them without paying
much attention to the actual data structures, this assumption no longer holds
for serialization. Indeed, we assume that an application will not try to
serialize everything in the right order using fine-grained
serialization combinators; instead, our definition of \lsf{iraw_data}, in the
array and map cases accommodates a union of two cases, allowing to mix unparsed raw
CBOR bytes (produced by \lsf{iparse_raw}) and recursive occurrences of
\lsf{iraw_data} built by the application. Then, we build a recursive serializer
for such raw CBOR data, where recursion is needed only for application-built
items, and user-controlled unparsed bytes are copied as is. Thus, the recursion
stack depth is entirely controlled by the application.

On the specification side, we define a recursive item serializer specification
and we prove it correct with respect to the corresponding parser. Since the
parser is injective, then the serializer is also injective.
The implementation combinator takes an \lsf{i:iraw_data}, an output byte array
and its length, serializing \lsf{i} into the output and returning the number of
bytes written, or 0 if the output buffer is too small.  We also implement a
function computing the size of the raw byte representation, without serializing
it.\iffull This takes as argument a
piece of raw CBOR data, and an upper bound (to protect against arithmetic
overflows), and returning the size of the byte representation, or 0 if it is
larger than the bound. The implementations of the two functions have the exact
same structure, apart from the actual output. \fi

\subsection{Specifying and Implementing the CBOR data model}

We refine the raw CBOR model of the previous section first to CBOR (ensuring that
maps have no duplicates) and then to Deterministically Encoded CBOR (ensuring
that map keys are sorted, and that integers are represented minimally). For
space reasons, we focus primarily on our main result that Deterministically
Encoded CBOR is non-malleable and can be validated in constant stack space.

\iffull
It is not enough to consider map key duplicates using mere equality on
raw CBOR data. The major complication comes from the fact that maps
can appear anywhere, including in keys; thus, to compare keys, we need
to know how to compare maps within those keys, and to even compare
those maps, we first need to know that those maps are themselves
valid. In this process, we need to forbid a map from having two entry
keys of equivalent representations, whether with integer
representations of different sizes,
or by the order of the map entry keys of the key itself.

To this end, on the specification side, we define two mutually recursive
predicates: for a raw CBOR data to be \emph{valid}, and for a pair of raw CBOR data to be \emph{equivalent}.
A piece of raw CBOR data $x$ is valid if, and only if, all of its
children data items (tagged payload, array items, map keys and values) are valid
and, if $x$ is a map, no two entries have equivalent map keys; and two pieces of
raw CBOR data are equivalent if, and only if, they are equal, or both valid and
of the same type, and, depending on their type, their integer values or simple
values are equal (regardless of their byte sizes), or their array items or tag
payloads are equivalent, or they are both key-value maps and their map entries
seen as dictionaries associate equivalent keys to equivalent values.

To typecheck these predicates in \fstar, we have to prove that the recursion is well-founded. To this end, we first define
the \emph{size} of a raw CBOR data by structural recursion: a raw CBOR
array has size 2 plus the sum of the sizes of its elements; a raw CBOR
map has size 2 plus the sum of the sizes of its entry keys and the
sizes of its entry values, a raw tagged CBOR data has size 2 plus the
size of the payload; any other raw CBOR data has size 1. Then, we
mutually define the validity and equivalence predicates by recursion
on the sum of the sizes of their arguments.

In spite of the recursive nature of its specification, the validity of
a piece of raw CBOR data $x$ can be implemented in a way similar to
the validator (or the jumper) for raw CBOR bytes, by maintaining a
counter for the number of remaining children items to visit. This loop
alone eliminates the need for a stack for the purpose of this
visit. Moreover, checking for map key duplicates can be performed by
two loops, one over the whole map, and one over the entries following
the current entry for which we need to check that there are no other
entries with an equivalent key. Thus, stack consumption only depends
on the stack consumption of equivalence checking.

However, equivalence checking in general cannot be performed in a
similar way, because of the order in map entries: checking the
equivalence of two maps requires a stack at least proportional to the
level of their map nesting.

\paragraph{No maps in map keys}
If a raw CBOR data has no maps in map keys, its validity can be
checked in constant stack space, because, equivalence  of map keys
themselves containing no maps can be checked in constant stack
space. This is enough for the COSE message layer, which mandates that
map keys can only contain text strings and
integers \cite[\S~1.5]{cose}. However, while this proves that validity
for this subset of CBOR can be checked in constant stack space, this
is not enough to define a formal data model for the whole CBOR.

\paragraph{Deterministically Encoded CBOR}
Fortunately, this limitation on map entry keys is not necessary,
thanks to the ``deterministic'' encoding of CBOR relying on minimal
integer byte sizes and map key ordering.
\fi

Given a total order on raw CBOR data, we first prove, by recursion on the
sizes of their input CBOR data, that a piece of CBOR data where all of its
integer byte sizes are minimal and all of its map keys are sorted with respect
to the strict order, is valid; and two such pieces of CBOR data that are
equivalent to each other are equal.
However, this is not enough to prove that this representation covers all
possible CBOR items. So, we prove, by induction on the size, that minimizing the
integer byte sizes of valid raw CBOR data headers (integer value, tag value, or
array or map entry count) $x$ yields a valid raw CBOR data equivalent to $x$.
Then, we prove that sorting the entries in a valid map where integers have
minimal representation results in a valid, equivalent map. Thus, on the
specification side, any valid raw CBOR data can be turned into such a
representation, by recursively minimizing all its integer byte representations
and sorting all its maps. Thus, we obtain the following:

\begin{theorem} \label{th:cbor}
Given a total strict order $<$ on raw CBOR data, the type $\mathsf{cbor}$ of
raw CBOR data with minimal integer byte representations and maps
sorted with respect to $<$ is a data model for CBOR, in the sense that
there is a bijection between $\mathsf{cbor}$ and the following \textrm{view} type:
\[
\begin{array}{rcl}
\textrm{view} & ::= & \textsf{Int} ~ (x \in [-2^{64}; 2^{64}-1]) \\
& | & \textsf{Simple} ~ (x \in [0, 23] \cup [32, 255]) \\
& | & \textsf{ByteString} ~ (n \in [0, 2^{64} - 1], x \in [0, 255]^n) \\
& | & \textsf{TextString} ~ (n \in [0, 2^{64} - 1], x \in [0, 255]^n \cap \textrm{UTF-8}) \\
& | & \textsf{Tagged} ~ (\textrm{tag} \in [0, 2^{64} - 1], v: \mathsf{cbor}) \\
& | & \textsf{Array} ~ (n \in [0, 2^{64} - 1], x \in \mathsf{cbor}^{n}) \\
& | & \textsf{Map} ~ (n \in [0, 2^{64} - 1], x: ~ (\mathsf{cbor} \xrightarrow{n} \mathsf{cbor}))
\end{array}
\]
and there is a function $\mathsf{size} : \mathsf{cbor} \rightarrow \mathbb
N$, such that a CBOR item
has always strictly larger size than any CBOR item appearing in its
view as its tagged payload, or an array or map entry.
\end{theorem}

\iffull
This \lsf{view} type is similar to but different than the mathematical
data model of Fig.~\ref{fig:cbor-raw-data}: the \lsf{view} type is not
recursive, it is rather meant as a way to case analyze on a CBOR data
item, where tag, array and map payloads are CBOR data items instead of
views. But this time, maps are true mathematical finite maps with no key
duplicates, and any integer byte sizes have disappeared.
\fi

The existence of the \emph{size} function with the property on the
view ensures that there are no \emph{cyclic} CBOR items (e.g. an item
that would appear itself in one of its tagged payloads, array or map
entries.)

We instantiate this theorem with the lexicographic ordering on the
byte representation of raw CBOR data with respect to the serializer
specification defined in \S~\ref{sec:cbor:raw}. Then, since that
serializer is injective, ``Deterministically Encoded CBOR'' is indeed
a unique representation of CBOR items.

On the verified implementation side, we implement a function checking
that raw CBOR bytes have minimal integer byte sizes and have their map
entries sorted with respect to a strict order. \iffull The structure of this
checker is similar to that of the jumper, where the stack consumption
only comes from the function that compares two map keys. \fi With the
lexicographic byte ordering, stack consumption is constant.

However, on the serialization side, instead of implementing a function that
would recursively sort map entries from valid raw CBOR bytes, we provide a
verified, defensive API that allows constructing CBOR items using C or Rust data
structures. As part of our verified API, we provide a function to create a CBOR
map from an array of pairs of CBOR items representing the map entries. This
function sorts the map entries in place without serializing them, thanks to the following
theorem reflecting the lexicographic byte representation order at the level of
the CBOR item view:

\begin{theorem} \label{th:cbor-compare}
Let $x_1$ and $x_2$ be two CBOR items of respective types \iffull(as defined
in Figure~\ref{fig:cbor-data-model})\fi $t_1$ and $t_2$. $x_1 < x_2$ with
respect to their deterministic byte representation if, and only if,
$t_1 < t_2$ , or $t_1 = t_2$ and one of the following holds:
\begin{enumerate}
\item they are both nonnegative integers, or simple values, and their values are ordered: $n_1 < n_2$
\item they are both negative integers, or simple values, and their values are counter-ordered: $-1-n_1 < -1-n_2$
\item they are both tagged items, and their tags $\textrm{tag}_1 < \textrm{tag}_2$, or $\textrm{tag}_1 = \textrm{tag}_2$ and their payloads $x'_1 < x'_2$
\item they are both array items, and their number of entries $n_1 < n_2$, or $n_1 = n_2$ and their lists of entries are lexicographically ordered with respect to $<$
\item they are both map items, and their number of entries $n_1 < n_2$, or $n_1 = n_2$ and their lists of entries, with the keys sorted wrt. $<$, are lexicographically ordered with respect to the lexicographic order on key-value pairs derived from $<$
\end{enumerate}
\end{theorem}

This theorem, leveraging big-endian encoding of integers of a given size,
justifies the use of the lexicographic byte ordering over the length-first byte
ordering defined in the previous version of the CBOR standard \cite{cbor-old}.

Our map creation function is defensive, in the sense that if it
encounters duplicate keys during sorting, it gracefully fails.

Then, since map entries are sorted in their data structure
representations, it is enough to reuse the raw CBOR data serializers
that we defined in \S~\ref{sec:cbor:raw}, using minimal byte
representations for integers, provided that user-controlled unparsed
CBOR bytes use the deterministic encoding. Indeed, we deem this
proviso necessary for security, because replacing bytes representing
valid CBOR data with their deterministic encoding would need to be
performed in depth-first fashion, thus requiring stack usage at least
proportional to the depth of the CBOR item.

Calling the serializer returns the byte size of the binary representation, or 0
if the output buffer is too small, as specified as the following separation
logic triple:
\begin{lstlisting}[language=fstar]
{ x $\uparrow$ v * b |-> s }  let n = iserialize x b
{ exists s'. x $\uparrow$ v * b |-> s' * ((n>0 <==> |serialize(v)| $\leq$ |s|) /\
    (n>0 ==> (n=|serialize v| /\ prefix n s'=serialize(v)))) }
\end{lstlisting}

We generate C and Rust serializers with the following signatures:
\begin{lstlisting}[language=pulse]
size_t iserialize(icbor x, uint8_t *output, size_t output_len);
fn iserialize <'a>(x: icbor <'a>, output: &'a mut [u8]) -> option__size_t
\end{lstlisting}

As such, one can use \evercbor directly from C or Rust, as a high-assurance,
full-featured CBOR library. Even among unverified implementations of CBOR,
QCBOR, a ``commercial-grade'' implementation, has long not
supported sorting of map keys until version 2.0, released in February 2025, and
which is still alpha as of April 2025, thus illustrating the intricacies of
implementing the deterministic encoding.

\paragraph{Limitations}
CBOR also allows representing floating-point numbers in
IEEE~754 \cite{ieee754} half-precision, single-precision and
double-precision formats. However, we do not support floating-point
numbers, due to lack of \fstar support, although formalizations of
floating-point values and their representations exist for other
theorem provers, such as Flocq for Coq \cite{flocq}. Moreover, the
statement of Theorem~\ref{th:cbor-compare} for floating-point values
would not be as simple as for integers, since the size prefix for
floating-point values in the CBOR binary encoding indicates precision
rather than magnitude.
CBOR also provides for definition of further types (long integers, dates, etc.)
as an interpretation of byte strings tagged with certain tags. Long integer
representations potentially overlap with the standard representations of 64-bit
integers, and a deterministic encoding allowing to conflate such representations
would actually further \emph{restrict} the space of valid byte representations.
We leave such extended data models to future work.

\section{\evercddl: Verified Parsers and Serializers for CDDL}
\label{sec:cddl}

Although CBOR, like JSON, was initially being designed as a schema-less binary
representation, most security-critical applications do not use CBOR as is, but
rather want to parse and serialize CBOR items following a schema of their
choice. To this end, in 2019, the IETF proposed CDDL (Concise Data Definition
Language, \cite{cddl}) as a schema language for CBOR.  While CDDL is still a
proposed standard, it has increasingly been used in other standards such as
COSE~\cite{cose}, DPE~\cite{dpe}, and SCITT~\cite{scitt}.

In this section, we introduce \evercddl, a formal model of CDDL in \fstar, and a
code generator that transforms a CDDL description to low-level types, parsers,
and serializers for CBOR items valid with respect to such a description.

\subsection{Syntax and Semantics} \label{sec:cddl:syntax}

A simplified syntax for CDDL descriptions is shown below:
\[
  \begin{array}{lrl}
 \mbox{type} &   t & ::= \theta \mathop{|} \lbrack a \rbrack \mathop{|} \{ g \} \mathop{|} t_1 / t_2 \\
 \mbox{base}  & \theta & ::= \bot \mathop{|} \ell \mathop{|} \texttt{any} \mathop{|} \texttt{int}\mathop{|} \texttt{uint}\mathop{|} \texttt{nint}\mathop{|} \texttt{tstr} \mathop{|} \texttt{bstr} \\
 \mbox{label} &   \ell & ::= n \in \left[-2^{64}, 2^{64} - 1\right) \mathop{|} s : \textrm{UTF-8} \\
 \mbox{array group} &   a & ::= t \mathop{|} a_1 \groupor a_2 \mathop{|} ?a \mathop{|} a_1, a_2 \mathop{|} *a \\
 \mbox{map group} & g & ::= t_k \Rightarrow t_v \mathop{|} \ell : t \mathop{|} g_1 \groupor g_2 \mathop{|} ?g \mathop{|} g_1, g_2 \mathop{|} *g \\
  \end{array}  
\]
We explain with an example: Two entities, a company and a nonprofit, want to
produce a record of their name, their status, and the names and salaries of its
employees, encoding as a CBOR item which would have the following JSON shape:
\begin{lstlisting}[language=json]
[ "ACME Corp.", "company", { "J.D.": 1842, "M.S.": 1729, "CEO": "J.D." } ]
[ "The Main St. Assoc.", "nonprofit", { "John S.": 0 } ]
\end{lstlisting}
Such CBOR items satisfy the following CDDL schema:
\begin{lstlisting}[language=cddl]
[ tstr, ("company" / "nonprofit"), { ? ("CEO": tstr), * (tstr => uint) } ]
\end{lstlisting}
matching an array of three CBOR items, the first being a text string for the
entity name, the second being either ``company'' or ``nonprofit'' as a text
string, and the third being a map containing an optional key-value entry with
key equal to the text string ``CEO'' and a text string value, and zero or more
key-value entries where keys are text strings for employee names, and values are
nonnegative integers for their salaries.

\paragraph{Types}
In \evercddl, we specify a CDDL type as a Boolean predicate on CBOR items:
taking the \lsf{cbor} type defining the data model of Theorem~\ref{th:cbor}, the
semantics of a CDDL type is a Boolean function $\mathsf{cbor} \rightarrow
\mathsf{bool}$. For each CDDL type construct, we define its semantics as a
predicate combinator. The standard dictates that the semantics of CDDL is with
respect to CBOR without presumption of deterministic encoding---so, one cannot,
assume, say, that map entries are ordered. Of course, CDDL can be and is used
with Deterministically Encoded CBOR for security-critical applications.

\paragraph{Array groups}
An array group is one of: a type to describe a single CBOR element
satisfying that type, an alternative choice of two array groups, an optional
array group, a concatenation of two array groups, or a finite repetition of an
array group (the Kleene star), which is interpreted in a greedy fashion,
similarly to Parsing Expression Grammars (PEG) \cite{peg}. \iffull Thus, if $a$
is an array group that consumes at least one CBOR item, then $*a, a$ will never
match, since the first $*a$ will have consumed all sublists matching $a$,
leaving none matching the second $a$. For a given array group $a$, the
CDDL array type with array group $a$ matches a CBOR item $x$ if, and only if,
$x$ is a CBOR array and $a$ consumes all of its entries.\fi
PEG semantics prescribe that the alternative is non-backtracking: $(a_1 \groupor
a_2), a$ is not equivalent to $(a_1, a) \groupor (a_2, a)$ in most cases, unless
$a$ always succeeds. \iffull Consider a CBOR item list $l$, and assume that $a_1$
succeeds on $l$ and returns remaining list $l'$. Then, if $a$ fails, the whole
array group fails, and the alternative $a_2$ will not be rechecked on $l$ again.
\fi
In \evercddl, we specify an array group as a function that takes a
list of CBOR items and returns a splitting pair of such a list,
consisting of the list of consumed items and the list of remaining
items; or \verb+None+ if the CBOR item does not match.

\paragraph{Map groups}
A map group is one of: an entry descriptor consisting of a type for the entry
key and a type for the entry value; or an optional map group; or a finite
repetition of a map group.
An entry descriptor can be equipped with a \emph{cut}, \iffull which is meant to
be the last possible matching rule for keys matching the key type,\fi in the
sense that if there is an entry whose key matches the key type but the value
does not match the value type, then the whole map fails to validate, regardless
of alternatives. For instance, the map $(18 \mapsto 21)$ matches $? (18
\Rightarrow 42)$, with no entry consumed; but it does not match $? (18 : 42)$,
because of the use of the cut `$:$' rather than `$\Rightarrow$'. \iffull Since
that cut is nested within an option $?$, its behavior is best described as an
``exception'' semantics. But since CDDL alternatives are not backtracking, there
is no way to ``catch'' such an exception in a CDDL schema.\fi
Just like array groups, a map group can be seen as a function taking a map,
potentially consuming some of its entries, and returning the map of unconsumed
entries, with concatenation being function composition.

\paragraph{Deterministic map groups} Unfortunately, not all map groups are
admissible in CDDL, since some of them can be ambiguous because CBOR map entries
are, in general, unordered. Consider the CBOR map $(18 \mapsto \text{``foo''});
(42 \mapsto \text{``bar''})$: the map group 
\verb+(uint => tstr)+ may match either of the two entries. 
Our semantics first defines the nondeterministic validity semantics of a map
group as a function that takes a finite CBOR map and returns either a \emph{set}
of possible consumed-remaining map pairs, or $\bot$ if a cut fails. Then, a map
group is \emph{deterministic} if, and only if, it returns $\bot$ or a singleton
set. We prove that, if $t_k$ and $t_v$ are CDDL types, then, even though $t_k
\Rightarrow t_v$ may be nondeterministic, $* (t_k \Rightarrow t_v)$ is always
deterministic, always succeeds, and consumes all map entries whose keys match
$t_k$ and values match $t_v$. We prove the following theorems.

\begin{theorem} \label{th:cddl:map-group-rewrite} 
If $t_1^k, t_1^v, t_2^k, t_2^v, \dots$ are CDDL types, and $o_1, o_2, \dots \in
\{ \Rightarrow, : \}$, then $* ((t_1^k \mathop{o_1} t_1^v) \groupor (t_2^k
\mathop{o_2} t_2^v) \groupor \dots)$ has the same validity semantics as $*(t_1^k
\mathop{o_1} t_1^v), *(t_2^k \mathop{o_2} t_2^v), \dots$. 
\end{theorem}

\begin{theorem} \label{th:cddl:det-map-group}
The subset of CDDL map groups defined as follows yields only
deterministic map groups:
\[
g ::= \ell \Rightarrow t \mathop{|} \ell : t \mathop{|} * (t_k \Rightarrow t_v) \mathop{|}  g_1 \groupor g_2 \mathop{|} ?g \mathop{|} g_1, g_2
\]
\end{theorem}
\newcommand\sem[1]{\ensuremath{\llbracket #1 \rrbracket}}

\paragraph{Type interpretation} Every CDDL type $t$ can be
interpreted as type in \fstar, $\sem{t}$. For instance,
$\sem{\mbox{\texttt{uint}}}$ is \lsf{U64.t}, the type of unsigned 64-bit
integers; $\sem{t_1/t_2}$ is 
\lsf{either $\sem{t_1}$ $\sem{t_2}$}, the disjoint union. Similarly, we turn
array or map group concatenation into a pair; the Kleene star for array groups
as a list; and the Kleene star for map groups as the type \lsf{Map.t key (list
value)}, finite associations, accommodating duplicate keys with unspecified key
ordering (subsequently, refined to forbid duplicates); and constant literals to
the \lsf{unit} high-level type. For our illustrative example, the high-level
type associated to an entity record is a tuple with a \lsf{string} for the
entity name; \lsf{either unit unit} corresponding to the company or nonprofit
alternative; \lsf{option(unit & string)} for the optional CEO field, and
\lsf{Map.t string (list U64.t)} for the employee name-salary table.

\paragraph{Ambiguity} The type interpretation exposes other challenges with
ambiguity as well. For instance, CDDL does not require alternatives to be
disjoint. Consider for instance the CDDL type $\texttt{uint} / \texttt{any}$.
However, if we naively serialize the value $\textsf{Inr}(\textsf{Int}(42))$,
which is the right-hand-side of the disjoint union type and parse it back, the
parser could return $\textsf{Inl}(42)$. As another example, consider the
following CDDL map group $(18 \Rightarrow \texttt{uint}), *(\texttt{uint}
\Rightarrow \texttt{any})$. If we try to serialize the high-level value $((() \mapsto [42]),
(18 \mapsto [21]))$, the serializer should fail because the two CBOR maps obtained
for each part of the concatenation will have non-disjoint domains, so it is
impossible to concatenate those CDDL maps. To identify and rule out such
ambiguities, we define an internal elaboration system for CDDL, which we
describe next.

\begin{figure}
  \begin{gather*}
  \frac{}{(t; (t_k \Rightarrow t_v)) \rightsquigarrow (t / t_k; (t_k \Rightarrow t_v))} \\
  \frac{}{(t; (\ell : t_v)) \rightsquigarrow (t / \ell; (\ell : t_v))} \\
  \frac{}{(t; ? (\ell : t_v)) \rightsquigarrow (t / \ell; ? (\ell : t_v))} \\
  \frac{(t / \ell; g_1) \rightsquigarrow (t_1; g_1') \qquad 
        (t / \ell; g_2) \rightsquigarrow (t_2; g_2')}
       {(t; ((\ell : t_v), g_1) \groupor g_2) \rightsquigarrow
        (t_1 \cap t_2; ((\ell : t_v), g_1') \groupor g_2')} \\
  \frac{(t; g_1) \rightsquigarrow (t_1; g_1') \qquad
        (t; g_2) \rightsquigarrow (t_2; g_2')}
       {(t; g_1 \groupor g_2) \rightsquigarrow (t_1 \cap t_2; g_1' \groupor g_2')} \\
  \frac{(t; g_1) \rightsquigarrow (t_1; g_1') \qquad
        (t_1; g_2) \rightsquigarrow (t_2; g_2')}
        {(t; g_1, g_2) \rightsquigarrow (t_2; g_1', g_2')} \\
  \frac{}{(t; * (t_k \Rightarrow t_v)) \rightsquigarrow (t; * ((t_k \backslash t) \Rightarrow t_v))} \\
  \end{gather*}
    \caption{Annotating map group tables with excluded sets of keys. For two types
  $t_1$, $t_2$, we compute an underapproximation $t_1 \cap t_2$ of their
  intersection.}
  \label{fig:cddl:annot-map-group}
  \end{figure}

\newcommand\elab{\ensuremath{\mbox{\textit{elab}}}}

\paragraph{Elaboration} Our elaboration of CDDL uses the extended syntax of
deterministic map or map groups (shown below), with decorations on its domain,
where
$* ((t_k \backslash t_{\textrm{rej}}) \Rightarrow t_v)$ is a table matching
entries whose keys match $t_k$ but not $t_{\textrm{rej}}$ and values match
$t_v$. 
\[
g ::= \ell \Rightarrow t \mathop{|} \ell : t 
  \mathop{|} * ((t_k \backslash t_{\textrm{rej}}) \Rightarrow t_v)
  \mathop{|}  g_1 \groupor g_2 
  \mathop{|} ?g
  \mathop{|} g_1, g_2
\]
Elaboration $\elab(t)$, is a partial function, proceeding in several steps.
First, we use Theorem~\ref{th:cddl:map-group-rewrite} to rewrite map groups into
a canonical form, and then check that map groups are all of the deterministic
form of Theorem~\ref{th:cddl:det-map-group}. If not, we reject the
specification.

Next, for each deterministic map group $g$, we annotate its tables with key type
specifications that should be rejected. To this end, we define the function $(t;
g) \rightsquigarrow (t'; g')$, defined in Figure~\ref{fig:cddl:annot-map-group},
saying that a map group $g$ applied to any map that has no keys matching $t$
behaves the same as $g'$, and if successful, the remaining map entries have no
keys matching $t'$. The rewrite rules are specified in priority order, so the
fourth rule takes precedence over the overlapping fifth rule. For a given map
descriptor $\{ g \}$, we rewrite $(\bot, g) \rightsquigarrow (t', g')$, and use
$g'$ as its elaborated form. Finally, we check the following properties,
rejecting $g'$ if any of them fail: (1) all alternatives must be disjoint; (2)
for any array groups $a_1$ and $a_2$, if $* a_1, * a_2, a_3$ appears, then $a_1$
and $a_2$ must be disjoint and $a_1$ and $a_3$ must be disjoint; and if $*a_1,
a_2$ appears, then $a_1$ and $a_2$ must be disjoint. (This is to avoid things
like $*a, a$, which we know will never match); and (3) for any map groups $g_1$
and $g_2$, if $g_1, g_2$ appears, then the footprints of $g_1$ (the types of all
keys appearing in $g_1$, minus the excluded keys $t_{\textrm{rej}}$ in $* ((t_k
\backslash t_{\textrm{rej}}) \Rightarrow t_v)$) and $g_2$ must be disjoint.

\begin{theorem}\label{th:cddl:elab-equiv} 
Given a CDDL type $t$, if $\elab(t)=t'$ is defined, then $t$ and $t'$ have
equivalent validating semantics: a CBOR item is valid for $t$ if and only if it
is valid for $t'$.
\end{theorem}

We also prove that elaborated types are unambiguous, though first we need to
introduce the semantics of CDDL parsers.

The elaboration described above is a simplification: indeed, we have
extended the implementation of \evercddl to annotate tables with
key-value footprints (instead of just keys), represented as Boolean
formulae where atoms are pairs of key-value types. This allows us to
support extensibility patterns such as
$?(18 \Rightarrow \texttt{uint}),
*(\texttt{uint} \Rightarrow \texttt{any})$, where the table
$*(\texttt{uint} \Rightarrow \texttt{any})$ can accept an entry with
key 18, provided its value is not an unsigned integer.

\paragraph{Parsing Semantics} \label{sec:cddl:parsing}
A main design goals of CDDL is to ``enable extraction of specific elements from
CBOR data for further processing'' \cite[\S~1]{cddl}, which basically means
parsing. The parsing specification of a CDDL type $t$ is a function taking a
CBOR item, item list or map valid with respect to $t$, and returning a value of
type $\sem{t}$. For instance, for \verb$uint$, the parser specification extracts
the integer value of a CBOR item an returns it as a \lsf{U64.t}. For $t_1 /
t_2$, the corresponding parser is $p(x) = \textsf{Inl}(p_1(x))$ if $x$ satisfies
$t_1$, and $\textsf{Inr}(p_2(x))$ otherwise, where $p_i$ is the parser for
$t_i$.

This brings us to our main theorem about the semantics of CDDL:

\begin{theorem}
\label{th:cddl:unambiguity} 
Given a CDDL type $t$, if $\elab(t)$ is defined, and $p$ is the parser
specification associated with $t$, then $p$ is injective; we can define a
\emph{serializability} function
$\sigma: u \rightarrow \textsf{bool}$, such that for any CBOR data $x$ valid
with respect to $t$, $\sigma(p(x))$ holds; and we can define a serializer
specification 
$s: (x: u \{ \sigma(x) \}) \rightarrow \textsf{cbor}$ such that for any
serializable high-level value $x$, $p(s(x)) = x$.
\end{theorem}

The serializability function $\sigma$ refines the type interpretation $\sem{t}$
to enforce constraints such as the absence of duplicate keys in maps.

\paragraph{Extensions and limitations} We have presented a simplified version of
what \evercddl supports. In particular, our implementation also supports integer
ranges, byte lengths, and UTF-8 strings.

We only support
non-recursive CDDL descriptions; while we investigated the formal semantics of
recursive CDDL descriptions, we ultimately deem them a security issue because
they would give rise to stack consumption proportional to the size of the input.
Standards such as COSE use recursion only up to a depth of 2 or 3, which is
easily supported by unrolling.

\subsection{Code Generation: Implementing Formatters for CDDL}

Once \evercddl elaborates and proves the unambiguity of a CDDL definition, it
generates implementation code in Pulse for types, validators, parsers, and serializers.

\paragraph{Validators} A validator for a CDDL type $t$ takes as argument a CBOR
item (obtained either from calling the \evercbor parser, or by constructing a
CBOR item using the \evercbor API) and returns a Boolean value, \texttt{true} if
and only if the CBOR item is valid with respect to $t$.
For CDDL array groups, the validator takes as argument a pointer to a CBOR array
iterator (the pointer is stack-allocated by the caller) and returns
\texttt{true} if and only if the array group succeeds, with the validator
advancing the iterator to consume the relevant array items.
For CDDL map groups, the validator takes as argument a CBOR item representing a
map, and a caller-allocated pointer to the number of map entries that have not
been consumed yet. Since the validators rely on the fact that \evercddl only
concatenates map groups with disjoint key domains, it is enough to count the
number of map entries left, and there is no need to precisely track which
entries have been consumed. Thus, validating a map does not require any heap
allocation, though incrementally validating the entries of a map may require
repeatedly scanning a prefix of already validated keys.

\paragraph{Parsers} The parser implementation for $t$ takes a CBOR item assumed
to be valid with respect to $t$, and returns a low-level representation $l :
\hat{\sem{t}}$ of the high-level value $h:\sem{t}$ returned by the parser
specification, similar to the definition of \ls$iparse_raw$ in \S\ref{sec:cbor}.
The difference is that \evercddl also generates the $l \uparrow v$ separation
logic predicate relating low-level and high-level values.
At the top-level, we combine the \evercbor validator and parser with the
\evercddl validator and parser, producing a function that takes as input a byte
array and its length, and returns a low-level representation of the result of
the CDDL parser specification and the remainder of the byte array\iffull past
the byte representation of the corresponding CBOR item\fi, or \verb+None+ if the
input bytes are not a valid representation of a CBOR object valid with respect
to $t$.

For a given array group $g$, the parser implementation takes as argument a
caller-allocated pointer to a CBOR array iterator assumed to be valid with
respect to $g$, and returns a low-level representation of the high-level value
returned by the parser specification. If $g$ is a Kleene star $g = *g'$, then,
similarly to \evercbor, we do not parse the full contents of the array. Rather,
we split the array iterator into two adjacent slices, the left-hand-side one
covering all array items consumed by $g$; then we return that iterator slice
along with a function pointer to the array parser for $g'$, leaving to the
application the responsibility of advancing that iterator to parse the array
elements. Map groups are similar, where for a table, we do not parse the full
contents of the map. Rather, we return a record value containing the CBOR map
and function pointers for the validator for the CDDL key type, the key exclusion
domain, and the value type, as well as parsers for the key and value types. The
validator function pointers are necessary since matching map entries are not
necessarily contiguous, contrary to arrays. We then provide a generic iterator
combinator to advance the map accordingly.

\paragraph{Serialization}
Contrary to parsing, we generate serializers that directly produce the
deterministic byte encoding of the CBOR item that is the result of the
serializer specification, rather than producing a CBOR data by allocating
intermediate \lsf{iraw_data} objects for use with the \evercbor API.
A serializer for $t$ takes as argument a low-level $l : \hat{\sem{t}}$, an
output byte array and its length, and returns the number of bytes written, or
0 if the output array is too small or if the high-level value is not
\emph{serializable} (e.g., it violates the serializability condition $\sigma$
from Theorem~\ref{th:cddl:unambiguity} with integer or simple value out of
bounds, invalid UTF-8 text bytes, etc.)

For the array descriptor and the map descriptor, the serializer first calls the
array group or map group serializer, then encodes the header with number of
entries written, then swaps the entries and the header. This is necessary for
the deterministic encoding if we want to traverse the input data at most once.
An alternative could be to traverse the input data twice, once to compute the
number of entries to write, and another one to serialize the entries. If we were
not using the deterministic CBOR encoding, we could always use 9 bytes to store
the number of entries (1 byte for the CBOR type, plus 8 bytes for the integer
encoding, see Fig.~\ref{fig:cbor-binary})

\iffull
For the Kleene star in array groups, we generate a serializer that
takes as argument either an array of low-level representations of
high-level values to serialize, or an array iterator that was the
result of a parser. Thus, the serializer can serialize the contents of
an array returned by another parser, provided the relations between
the low-level array item representation and the high-level value match
between the parser and the serializer. To this end, we strive to make
the relation depend as little as possible on the parser specification.
\fi

For map groups, we generate a serializer that takes an output buffer already
containing some map entries sorted with respect to the lexicographic byte order,
and inserts serialized map entries into it, using sorted insert: for each entry
to insert, the serializer first writes it next to the existing output map, then
it scans the output map, comparing keys to determine where to insert the new
entry, then it swaps the new entry with the tail of the output map that follows
the insertion point. In doing so, it can detect that an entry with the same key
already exists in the output map. In that case, the serializer gracefully fails.
This is interesting especially for tables: this check on serialized output maps
is a sound way to check that the \emph{input} map has no duplicates. \iffull
From the verification point of view, the high-level datatype is neither a map
(because the serializer does not need to assume that the input map has no
duplicate keys) nor a list of entries (because the serializer does not need to
know about the order of entries), but a map between keys and lists of values:
this way, keys are not ordered, but, for a given key, the length of the list of
values equals the number of occurrences of the key in the ``map''. \fi

\section{Performance Benchmarks \& Verified Applications}
\label{sec:experiments}

In this section, we report on experiments using our verified tools, with both
quantitative and qualitative results. It's worth noting that our verified code
worked correctly the first time on all experiments. On an Intel Xeon E5-2680 v4
with 56 cores (1.2 GHz), using 24 cores, \pulseparse (650 lines for
\lsf{parse_rec} and its proofs + 700 lines for its implementation + 6400 lines
for all the Pulse combinators) verifies in 6 minutes, 
\evercbor (6k lines of spec + 26k lines of implementation and proofs) verifies in 10 minutes 
and extracts and compiles to both C and Rust in 1 minute.
Finally, \evercddl (6k lines of spec + 23k lines of implementation and proofs)
verifies in 0.5 hour.

Although we generate both C and Rust code, we focus on evaluating the
performance of the generated C code, unless explicitly stated
otherwise.

\subsection{Synthetic Benchmarks}

We evaluate \evercbor and \evercddl on several synthetic benchmarks, and show
that its performance is comparable with that of existing (unverified) libraries,
namely QCBOR and TinyCBOR, even though we have not had the time to implement any
optimizations after these initial benchmarking results.

Our first benchmark considers a record type with 8 fields of type \texttt{uint},
with results in the first line of Table~\ref{tab:perf}. From a CDDL description
(elided), \evercddl generates a struct type and parsers and serializers for it.
The QCBOR and TinyCBOR libraries do not provide CDDL functionality, so we write
C functions translating between the CBOR representation and a flat C structure.
The performance of our validator and parser is between QCBOR and TinyCBOR.  We
believe we can close the gap to QCBOR since, by default, \evercddl returns
parsed records as structures on the stack, rather than using out parameters to fill an
existing object---it should be straightforward to add support for this and close
the gap to QCBOR. For serialization, our code is slower than QCBOR and
TinyCBOR because we serialize in the deterministic encoding,
perhaps shuffling elements. %

Our second benchmark involves large maps. The relevant CDDL description is
simply $\texttt{map} = { * (\texttt{uint} => \texttt{uint}) }$. The benchmark
consists of a map with $N = 8000$ entries with random keys and values, which is
then looked up $K = 1000$ times with random keys, which may or may not be
present.
We begin from a serialized map in the deterministic encoding. For \evercddl, we
first validate this bitstring, which checks that the keys are in order,
obtaining an iterator. To look up a value, we construct a CBOR object from our
desired key, and call an \evercbor function to look it up in the map.
The QCBOR API offers a function for map lookup, so we use it.
For TinyCBOR, we iterate through the map comparing keys.
Here, \evercbor is faster than the other two libraries, for two main reasons.
One, given that we validated the map, we know the keys are in order and can
therefore stop early safely (we also stop early with TinyCBOR). Second,
importantly, since we know the object is deterministically encoded, we can
compare the serialized representation of keys directly, byte-for-byte, instead
of having to parse back the keys in the map.

Our third benchmark involves nested arrays, generated by
the CDDL description
$\texttt{arr} = [* \texttt{subarr} ]; \texttt{subarr} = [* \texttt{uint}]$.
By running \evercddl on this description, we generate a C type
for an \texttt{arr}, alongside a parser and serializer for it.
We measure the time it takes to serialize and parse an array of $N = 10^4$ where
every subarray also contains $N$ elements all set to zero, for a total of $10^8$
elements.
The CBOR object involved is roughly 100MB. For \evercddl, we generate a
structure in memory and call the serializer.
The QCBOR library, instead, provides a streaming API where elements are output
or parsed one at a time.
We include the setup time in the measurements for \evercddl, for a conservative
comparison.
\evercddl is less performant than the other libraries,
there are a few non-fundamental reasons for this.
For example, when writing each integer into the buffer,
there is a size check performed. This check involves constructing
the CBOR object (of a single integer) to be written, computing
its size, and checking that the remaining space is at least that.
This computation is rather wasteful and hard to optimize by the C compiler.
Specializing it manually, replacing the size of the CBOR integer by the constant
1, provides a 20\% performance improvement. We are confident
we can adjust our verified implementations to generate specialized
sizes to attain this speedup.
 
For parsing, there is a design difference between the APIs provided by \evercddl
and the other two libraries. \evercddl requires the buffer to be validated
before any data can be read from it, which incurs one full pass of the ~100MB
buffer. Once validated, the client code can use the iterators to walk the
object, without incurring copies, and extract the integers in it. The other
libraries provide streaming APIs that can simply walk the buffer and read the
integers on demand, avoiding the need for the initial pass, but allowing to
partially read a corrupted object.
For security-sensitive applications, we argue that a validation pass
should be performed in all cases, hence our benchmark for QCBOR and TinyCBOR
also include one such pass.
For validation, all three libraries perform similarly. However, our parsing is
slower, because the \evercddl iterator for the outer array is not related at all
to that of the inner array. Once the inner iterator reaches the end, and we want
to advance to next subarray, the outer iterator has to walk the buffer again to
find the new offset. Our current iterator API does not expose this fact, mainly because
it treats unparsed
and application-built data uniformly.
 
All in all, while there are some improvements to be made, our benchmarks show
performance close to a state-of-the-art unverified library, although our code
parses to and from application-level types with a verified, defensive
implementation.

\begin{table}
  \begin{tabular}{l|c|c|c|c|c|c|}
      & \multicolumn{2}{c|}{\evercddl} & \multicolumn{2}{c|}{QCBOR} & \multicolumn{2}{c|}{TinyCBOR} \\
      \hline
      & V/P & S & V/P & S & V/P & S \\
      \hline
  Rec ($\mu s$) & 3.33 & .57
      & 1.91 & .23 & 3.78 & .29 \\
       \hline
  Map ($\mu s$) & \multicolumn{2}{c|}{138} & \multicolumn{2}{c|}{282} & \multicolumn{2}{c|}{306} \\
       \hline
  Arr (s)& 2.67/4.92 & 2.06 & 2.92/2.91 & 0.75 & 2.68/2.68 & 1.23 \\
      \hline

  \end{tabular}
  \caption{Synthetic benchmarks for \evercddl, QCBOR and TinyCBOR.
    Values are time (for Rec, for \textbf{V}alidation plus \textbf{P}arsing,
    or \textbf{S}erialization),
    lookup time (for Map),
    or time (for Arr).\iffull We distinguish validation
    from parsing in Arr, since iteration is involved.\fi}
  \label{tab:perf}
\end{table}

\subsection{Verified Applications: COSE \& DPE}
\label{sec:cose}

COSE is an Internet standard for signing and encryption of CBOR objects,
initially for securing the transport of IoT messages, though it is also used
today in non-IoT settings.
For signing, COSE defines a signature envelope message format
containing a signature structure. The signature structure is encoded
using Deterministically Encoded CBOR to make sure its byte
representation is unique; then, it is authenticated using
cryptographic hashing algorithms.

In the COSE standard, the message formats are described normatively in prose,
but they are accompanied with a non-normative CDDL description. We found that
the latter does not reflect the normative prose on two aspects, namely the
constraint that keys 5 and 6 must not appear together in \lsf{Generic\_Headers},
and an erroneously backtracking (non-PEG) interpretation of \lsf{?} in
\lsf{Sig\_structure}. So, we fixed the CDDL description accordingly.

With \evercddl, we support signature and verification formats with a single
(\textsf{COSE\_Sign1}) or multiple signers (\textsf{COSE\_Sign}), as well as
some cryptographic key object formats.

A notable limitation of our implementation of COSE is that the parser only
supports deterministic CBOR. Hence our implementation will reject COSE messages
that are not deterministically encoded. This is not a problem when serializing
messages since it is always allowed to write CBOR deterministically.

\paragraph{Evaluation}

\begin{table}
  \begin{tabular}{l|c|c}
    & C API \& OpenSSL & Pulse API \& \haclstar \\
    \hline
    COSE\_sign & 39.0 $\mu s$/iter & 53.3 $\mu s$/iter \\
    COSE\_verify & 99.6 $\mu s$/iter & 58.2 $\mu s$/iter \\
    \hline
    Ed25519\_sign & 36.8 $\mu s$/iter & 51.9 $\mu s$/iter \\
    Ed25519\_verify & 96.7 $\mu s$/iter & 57.3 $\mu s$/iter \\
    \hline
    parse(Sign1) & \multicolumn{2}{c}{2.4 $\mu s$/iter} \\
    ser(Sign1) & \multicolumn{2}{c}{1.0 $\mu s$/iter} \\
    ser(Sig\_structure) & \multicolumn{2}{c}{1.0 $\mu s$/iter} \\
  \end{tabular}
  \caption{Benchmarking results of our \evercddl-based COSE signature implementation.
    We sign and verify a message with an 896 byte long payload using Ed25519.
    The benchmarks were compiled with clang 19.1.7 (\texttt{-O3}) and run on an Intel Xeon W-2255 CPU.}
  \label{tbl:cosebench}
\end{table}

The \fstar file generated by \evercddl for the COSE specification takes 5
minutes to verify on a single core; extracting to C using Karamel takes another
23 seconds.
To evaluate interoperability and benchmark the performance
of the \evercddl-generated code,
we implement a small signature generation and verification tool
(limited to a single signer, Ed25519 algorithm, empty AAD, fixed headers)
in two versions:
both an unverified one using the \evercddl-generated C API and OpenSSL,
as well as a verified one using the Pulse API
and using the \haclstar library for cryptographic operations.
The benchmarking results in Table~\ref{tbl:cosebench}
show that the cryptographic primitives take up the majority of the runtime,
in both the verified and unverified versions.\footnote{
  We were surprised that OpenSSL signature verification is three times slower than signing
  and also slower than the fully verified \haclstar implementation.
  The \texttt{t\_cose} library however exhibits the same phenomenon (showing nearly identical performance),
  which is perhaps a sign of the complexity of using the OpenSSL API.
}

To give a flavor of the automatically generated C API, let us look at the CDDL
schema for \verb+COSE_Key_OKP+. This type specializes \verb+COSE_Key+ in the
COSE RFC to OKP keys; specializing the type makes \evercddl parse the fields for
the public key (-2) and private key (-4), and we do not need to go through the
map manually.

\begin{lstlisting}[language=cddl]
COSE_Key_OKP = { 1:1, -1:int/tstr, ?-2:bstr, ?-4:bstr, *label=>values }
\end{lstlisting}

On the C side, we get a structure and two functions, for serialization and
parsing.\iffull\footnote{We shorten namespaces in the generated C code.}\fi The structure has four fields: three for
explicitly specified data fields (-1, -2, and -4) and one for the map at the
end. The entry \verb+1:1+ does not correspond to a field in the C structure,
since \evercddl knows it just has the value 1. Types like \verb+option___bstr+
are created by Karamel using monomorphization.

\begin{lstlisting}[language=C]
typedef struct {
  label intkeyneg1; option___bstr intkeyneg2; option___bstr intkeyneg4;
  either___slice___map_iterator_t _x0;  } COSE_Key_OKP;
size_t serialize_COSE_Key_OKP(COSE_Key_OKP c, slice___uint8 out);
option___COSE_Key_OKP___slice_uint8
  validate_and_parse_COSE_Key_OKP(slice___uint8 s);
\end{lstlisting}

The Pulse API generated by \evercddl is expressive enough to state a precise
functional correctness specification. For signature verification, we define a
predicate relating a valid signature message with its payload, where \lsf{vmsg}
is the specification-level struct carrying the signed bytes,
while \lsf{tbs} is the bytes to be signed:

\begin{lstlisting}
let good_sig pubkey msg payload = exists vmsg tbs.
  parses_from bundle_COSE_Sign1_Tagged.b_spec vmsg msg /\
  vmsg.payload == Inl payload /\ length vmsg.sig == 64 /\
  to_be_signed_spec vmsg.protected payload tbs /\
  spec_ed25519_verify pubkey tbs vmsg.sig
\end{lstlisting}

The \ls{verify} function then takes fractional (shared) permissions
to the public key and (serialized) message,
and returns an optional slice for the payload.
The postcondition ensures that any payload returned by \ls{verify}
is signed by the given public key.

\begin{lstlisting}[language=fstar]
{ pubkey |->(r) vk * msg |->(p) vm } let payload = verify pubkey msg
{ pubkey |->(r) vk * (match payload with | None -> msg |->(p) vm
  | Some r -> exists vp q. (r |->(q) vp) >* (msg |->(p) vm) * good_sig vk vm vq) }
\end{lstlisting}

Similarly, signature generates guarantees that the output buffer
is a well-formed \verb+COSE_Sign1_Tagged+ object whose signature field
is a valid signature of the appropriate \verb+Sig_structure+.

\paragraph{DPE} We also specify the CDDL API for DPE, a secure boot protocol,
with functional correctness proofs, covering six different message types for the
four main functions on the DPE interface. Appendix~\ref{appendix:dpe} provides
some more information, though the main takeaway is similar to what we report for
COSE: \evercddl specifications are precise enough to express full functional
correctness of application code manipulating messages in a given CDDL schema.

\section{Related Work \& Conclusions}
\label{sec:related}
\newcommand\citepos[1]{\citeauthor{#1}'s\ (\citeyear{#1})}

\citet{bratus17curing} provide a useful perspective on the important of parsing
for software security, including guidelines for how to securely handle
attacker-controlled input. 

The most closely related line of work to ours is EverParse~\cite{everparse},
which we have discussed throughout the paper, since we reuse some of their
purely functional specification combinators. Many others have looked at purely
functional verified parsers and serializers. \citet{blaudeau20packrat} build a
verified packrat parser for parsing expression grammars
(PEGs)~\citep{ford02packrat,ford04peg} in the PVS proof assistant~\citep{pvs96},
while~\citep{parsley20} supports PEGs with constraints.
\citet{lasser19itp} build a verified implementation of an LL(1) parser generator
and ~\citet{costar21} verified an implementation of the ALL(*) parsing
algorithm, both in the Coq proof assistant. \citet{asn1star} use EverParse's
specification combinators to formalize ASN.1 DER~\citep{asn1}, a widely used
data formatting standard with goals similar to CBOR and CDDL, proving that ASN.1
DER is non-malleable. They use an ad hoc approach to formalizing the recursion
present in ASN.1, rather than our general purpose \lsf{parse_rec} combinator
with constant-stack-space validation. Ni et al. extract their specifications to
OCaml code, rather than going to fully low-level code in C, as we do.
Similarly, \citet{debnath24armor} also focus on ASN.1 and formalize it in Agda,
producing functional Haskell code for X.509 certificate chain validation.
\citet{delaware19narcissus} implement a combinator library for verified parsers
and serializers for binary formats in Coq, but they focus on producing purely
functional programs in OCaml, rather than zero-copy, low-level code. They also
do not prove non-malleability of formats.

Others have also looked at tools for low-level parsing and serializing.
\citepos{nailosdi2015} Nail is a DSL for writing low-level applications while
processing a given data format. It produces C code, but does not aim at
verification.~\citepos{daedalus24} Daedalus is a DSL with parser combinators
targeting both Haskell and C++, aiming to produce memory safe C++, but without
formal proof. Daedalus has been used at scale, including to generate parsers for
the PDF document standard. Daedalus does not support serialization. Recent
unpublished work describes a tool called
Vest ({\small\url{https://github.com/secure-foundations/vest}}) a parser and
serializer generator embedded in Verus~\cite{verus24sosp}, a dialect of Rust
aimed at verification. Vest's use of linear types in Rust is similar in spirit
to our use of separation logic. However, Vest does not support recursive formats
which are required to formalize languages like CBOR. \pulseparse is not tied
to Rust, and Pulse can in general be used to produce verified C code or
verified, safe Rust code, though support for the latter is not fully
complete.

\paragraph{Conclusions} In summary, we have presented a new approach to secure,
low-level formatting with foundations in separation logic. We have used this
foundation to develop a comprehensive, mechanized formalization of CBOR and
CDDL, two data formatting standards of significant stature in security-related
protocols, and applied our tools, including  formally verified libraries and
code generators, to a variety of other standards grounded in CBOR. We hope our
open-source tools will help others build systems that process these binary
formats correctly and securely.

\newpage

\bibliographystyle{ACM-Reference-Format}
\bibliography{paper}


\begin{thebibliography}{43}


\ifx \showCODEN    \undefined \def \showCODEN     #1{\unskip}     \fi
\ifx \showISBNx    \undefined \def \showISBNx     #1{\unskip}     \fi
\ifx \showISBNxiii \undefined \def \showISBNxiii  #1{\unskip}     \fi
\ifx \showISSN     \undefined \def \showISSN      #1{\unskip}     \fi
\ifx \showLCCN     \undefined \def \showLCCN      #1{\unskip}     \fi
\ifx \shownote     \undefined \def \shownote      #1{#1}          \fi
\ifx \showarticletitle \undefined \def \showarticletitle #1{#1}   \fi
\ifx \showURL      \undefined \def \showURL       {\relax}        \fi
\providecommand\bibfield[2]{#2}
\providecommand\bibinfo[2]{#2}
\providecommand\natexlab[1]{#1}
\providecommand\showeprint[2][]{arXiv:#2}

\bibitem[Bangert and Zeldovich(2015)]%
        {nailosdi2015}
\bibfield{author}{\bibinfo{person}{Julian Bangert} {and}
  \bibinfo{person}{Nickolai Zeldovich}.} \bibinfo{year}{2015}\natexlab{}.
\newblock \showarticletitle{Nail: {A} Practical Tool for Parsing and Generating
  Data Formats}.
\newblock \bibinfo{journal}{\emph{login Usenix Mag.}} \bibinfo{volume}{40},
  \bibinfo{number}{1} (\bibinfo{year}{2015}).
\newblock
\urldef\tempurl%
\url{https://www.usenix.org/publications/login/feb15/bangert}
\showURL{%
\tempurl}


\bibitem[Birkholz et~al\mbox{.}(2025)]%
        {scitt}
\bibfield{author}{\bibinfo{person}{H. Birkholz}, \bibinfo{person}{A.
  Delignat-Lavaud}, \bibinfo{person}{C. Fournet}, \bibinfo{person}{Y.
  Deshpande}, {and} \bibinfo{person}{S. Lasker}.}
  \bibinfo{year}{2025}\natexlab{}.
\newblock \bibinfo{title}{{ An Architecture for Trustworthy and Transparent
  Digital Supply Chains (SCITT)}}.
\newblock \bibinfo{howpublished}{draft-ietf-scitt-architecture-11}.
\newblock
\urldef\tempurl%
\url{https://www.ietf.org/archive/id/draft-ietf-scitt-architecture-11.txt}
\showURL{%
\tempurl}


\bibitem[Birkholz et~al\mbox{.}(2019)]%
        {cddl}
\bibfield{author}{\bibinfo{person}{Henk Birkholz}, \bibinfo{person}{Christoph
  Vigano}, {and} \bibinfo{person}{Carsten Bormann}.}
  \bibinfo{year}{2019}\natexlab{}.
\newblock \bibinfo{title}{{Concise Data Definition Language (CDDL): A
  Notational Convention to Express Concise Binary Object Representation (CBOR)
  and JSON Data Structures}}.
\newblock \bibinfo{howpublished}{IETF RFC 8610}.
\newblock
\href{https://doi.org/10.17487/RFC8610}{doi:\nolinkurl{10.17487/RFC8610}}


\bibitem[Blaudeau and Shankar(2020)]%
        {blaudeau20packrat}
\bibfield{author}{\bibinfo{person}{Clement Blaudeau} {and}
  \bibinfo{person}{Natarajan Shankar}.} \bibinfo{year}{2020}\natexlab{}.
\newblock \showarticletitle{A Verified Packrat Parser Interpreter for Parsing
  Expression Grammars}. In \bibinfo{booktitle}{\emph{Proceedings of the 9th ACM
  SIGPLAN International Conference on Certified Programs and Proofs}} (New
  Orleans, LA, USA) \emph{(\bibinfo{series}{CPP 2020})}.
  \bibinfo{publisher}{Association for Computing Machinery},
  \bibinfo{address}{New York, NY, USA}, \bibinfo{pages}{3–17}.
\newblock
\showISBNx{9781450370974}
\href{https://doi.org/10.1145/3372885.3373836}{doi:\nolinkurl{10.1145/3372885.3373836}}


\bibitem[Boldo and Melquiond(2011)]%
        {flocq}
\bibfield{author}{\bibinfo{person}{Sylvie Boldo} {and}
  \bibinfo{person}{Guillaume Melquiond}.} \bibinfo{year}{2011}\natexlab{}.
\newblock \showarticletitle{Flocq: A Unified Library for Proving Floating-Point
  Algorithms in Coq}. In \bibinfo{booktitle}{\emph{Proceedings of the 2011 IEEE
  20th Symposium on Computer Arithmetic}} \emph{(\bibinfo{series}{ARITH '11})}.
  \bibinfo{publisher}{IEEE Computer Society}, \bibinfo{address}{USA},
  \bibinfo{pages}{243–252}.
\newblock
\showISBNx{9780769543185}
\href{https://doi.org/10.1109/ARITH.2011.40}{doi:\nolinkurl{10.1109/ARITH.2011.40}}


\bibitem[Bormann and Hoffman(2013)]%
        {cbor-old}
\bibfield{author}{\bibinfo{person}{Carsten Bormann} {and}
  \bibinfo{person}{Paul~E. Hoffman}.} \bibinfo{year}{2013}\natexlab{}.
\newblock \bibinfo{title}{{Concise Binary Object Representation (CBOR)}}.
\newblock \bibinfo{howpublished}{IETF RFC 7049}.
\newblock
\href{https://doi.org/10.17487/RFC7049}{doi:\nolinkurl{10.17487/RFC7049}}


\bibitem[Bormann and Hoffman(2020)]%
        {cbor}
\bibfield{author}{\bibinfo{person}{Carsten Bormann} {and}
  \bibinfo{person}{Paul~E. Hoffman}.} \bibinfo{year}{2020}\natexlab{}.
\newblock \bibinfo{title}{{Concise Binary Object Representation (CBOR)}}.
\newblock \bibinfo{howpublished}{IETF RFC 8949}.
\newblock
\href{https://doi.org/10.17487/RFC8949}{doi:\nolinkurl{10.17487/RFC8949}}


\bibitem[Boyland(2003)]%
        {boyland03frac}
\bibfield{author}{\bibinfo{person}{John Boyland}.}
  \bibinfo{year}{2003}\natexlab{}.
\newblock \showarticletitle{Checking Interference with Fractional Permissions}.
  In \bibinfo{booktitle}{\emph{Static Analysis}},
  \bibfield{editor}{\bibinfo{person}{Radhia Cousot}} (Ed.).
  \bibinfo{publisher}{Springer Berlin Heidelberg}, \bibinfo{address}{Berlin,
  Heidelberg}, \bibinfo{pages}{55--72}.
\newblock
\showISBNx{978-3-540-44898-3}


\bibitem[Bratus et~al\mbox{.}(2017)]%
        {bratus17curing}
\bibfield{author}{\bibinfo{person}{Sergey Bratus}, \bibinfo{person}{Lars
  Hermerschmidt}, \bibinfo{person}{Sven~M. Hallberg},
  \bibinfo{person}{Michael~E. Locasto}, \bibinfo{person}{Falcon Momot},
  \bibinfo{person}{Meredith~L. Patterson}, {and} \bibinfo{person}{Anna
  Shubina}.} \bibinfo{year}{2017}\natexlab{}.
\newblock \showarticletitle{Curing the Vulnerable Parser: Design Patterns for
  Secure Input Handling}.
\newblock \bibinfo{journal}{\emph{login Usenix Mag.}} \bibinfo{volume}{42},
  \bibinfo{number}{1} (\bibinfo{year}{2017}).
\newblock
\urldef\tempurl%
\url{https://www.usenix.org/publications/login/spring2017/bratus}
\showURL{%
\tempurl}


\bibitem[B{\"u}ttner and Gruschka(2023)]%
        {fido-proof}
\bibfield{author}{\bibinfo{person}{Andre B{\"u}ttner} {and}
  \bibinfo{person}{Nils Gruschka}.} \bibinfo{year}{2023}\natexlab{}.
\newblock \showarticletitle{Protecting FIDO Extensions Against
  Man-in-the-Middle Attacks}. In \bibinfo{booktitle}{\emph{Emerging
  Technologies for Authorization and Authentication}},
  \bibfield{editor}{\bibinfo{person}{Andrea Saracino} {and}
  \bibinfo{person}{Paolo Mori}} (Eds.). \bibinfo{publisher}{Springer Nature
  Switzerland}, \bibinfo{address}{Cham}, \bibinfo{pages}{70--87}.
\newblock
\showISBNx{978-3-031-25467-3}


\bibitem[Debnath et~al\mbox{.}(2024)]%
        {debnath24armor}
\bibfield{author}{\bibinfo{person}{Joyanta Debnath}, \bibinfo{person}{Christa
  Jenkins}, \bibinfo{person}{Yuteng Sun}, \bibinfo{person}{Sze~Yiu Chau}, {and}
  \bibinfo{person}{Omar Chowdhury}.} \bibinfo{year}{2024}\natexlab{}.
\newblock \showarticletitle{ARMOR: A Formally Verified Implementation of X.509
  Certificate Chain Validation}. In \bibinfo{booktitle}{\emph{2024 IEEE
  Symposium on Security and Privacy (SP)}}. \bibinfo{pages}{1462--1480}.
\newblock
\href{https://doi.org/10.1109/SP54263.2024.00220}{doi:\nolinkurl{10.1109/SP54263.2024.00220}}


\bibitem[Decker and Wattenhofer(2014)]%
        {decker2014bitcoin}
\bibfield{author}{\bibinfo{person}{Christian Decker} {and}
  \bibinfo{person}{Roger Wattenhofer}.} \bibinfo{year}{2014}\natexlab{}.
\newblock \showarticletitle{Bitcoin transaction malleability and {MtGox}}. In
  \bibinfo{booktitle}{\emph{European Symposium on Research in Computer
  Security}}. Springer, \bibinfo{pages}{313--326}.
\newblock


\bibitem[Delaware et~al\mbox{.}(2019)]%
        {delaware19narcissus}
\bibfield{author}{\bibinfo{person}{Benjamin Delaware}, \bibinfo{person}{Sorawit
  Suriyakarn}, \bibinfo{person}{Cl{\'{e}}ment Pit{-}Claudel},
  \bibinfo{person}{Qianchuan Ye}, {and} \bibinfo{person}{Adam Chlipala}.}
  \bibinfo{year}{2019}\natexlab{}.
\newblock \showarticletitle{Narcissus: correct-by-construction derivation of
  decoders and encoders from binary formats}.
\newblock \bibinfo{journal}{\emph{Proc. {ACM} Program. Lang.}}
  \bibinfo{volume}{3}, \bibinfo{number}{{ICFP}} (\bibinfo{year}{2019}),
  \bibinfo{pages}{82:1--82:29}.
\newblock
\href{https://doi.org/10.1145/3341686}{doi:\nolinkurl{10.1145/3341686}}


\bibitem[Diatchki et~al\mbox{.}(2024)]%
        {daedalus24}
\bibfield{author}{\bibinfo{person}{Iavor~S. Diatchki}, \bibinfo{person}{Mike
  Dodds}, \bibinfo{person}{Harrison Goldstein}, \bibinfo{person}{Bill Harris},
  \bibinfo{person}{David~A. Holland}, \bibinfo{person}{Benoit Razet},
  \bibinfo{person}{Cole Schlesinger}, {and} \bibinfo{person}{Simon Winwood}.}
  \bibinfo{year}{2024}\natexlab{}.
\newblock \showarticletitle{Daedalus: Safer Document Parsing}.
\newblock  \bibinfo{volume}{8}, \bibinfo{number}{PLDI}, Article
  \bibinfo{articleno}{180} (\bibinfo{date}{June} \bibinfo{year}{2024}),
  \bibinfo{numpages}{25}~pages.
\newblock
\href{https://doi.org/10.1145/3656410}{doi:\nolinkurl{10.1145/3656410}}


\bibitem[Ebner et~al\mbox{.}(2025)]%
        {pulsecore}
\bibfield{author}{\bibinfo{person}{Gabriel Ebner}, \bibinfo{person}{Guido
  Mart\'{i}nez}, \bibinfo{person}{Aseem Rastogi}, \bibinfo{person}{Thibault
  Dardinier}, \bibinfo{person}{Megan Frisella}, \bibinfo{person}{Tahina
  Ramananandro}, {and} \bibinfo{person}{Nikhil Swamy}.}
  \bibinfo{year}{2025}\natexlab{}.
\newblock \showarticletitle{PulseCore: An Impredicative Concurrent Separation
  Logic for Dependently-Typed Programs}. In \bibinfo{booktitle}{\emph{46th ACM
  SIGPLAN Conference on Programming Language Design and Implementation
  (PLDI)}}. \bibinfo{publisher}{ACM}.
\newblock
\newblock
\shownote{(accepted for publication, to appear)}.


\bibitem[{European Union eHealth Network}(2021)]%
        {cddlcovid}
\bibfield{author}{\bibinfo{person}{{European Union eHealth Network}}.}
  \bibinfo{year}{2021}\natexlab{}.
\newblock \bibinfo{title}{European Union Digital COVID Certificate (EUDCC)
  Electronic Health Certificates Specification}.
\newblock
  \bibinfo{howpublished}{\url{https://github.com/ehn-dcc-development/eu-dcc-hcert-spec}}.
\newblock


\bibitem[Finney(2006)]%
        {finney2006bleichenbacher}
\bibfield{author}{\bibinfo{person}{Hal Finney}.}
  \bibinfo{year}{2006}\natexlab{}.
\newblock \showarticletitle{Bleichenbacher's RSA signature forgery based on
  implementation error}.
\newblock
  \bibinfo{journal}{\emph{\url{https://mailarchive.ietf.org/arch/msg/openpgp/5rnE9ZRN1AokBVj3VqblGlP63QE/}}}
  (\bibinfo{year}{2006}).
\newblock


\bibitem[Ford(2002)]%
        {ford02packrat}
\bibfield{author}{\bibinfo{person}{Bryan Ford}.}
  \bibinfo{year}{2002}\natexlab{}.
\newblock \showarticletitle{Packrat parsing: : simple, powerful, lazy, linear
  time, functional pearl}. In \bibinfo{booktitle}{\emph{Proceedings of the
  Seventh {ACM} {SIGPLAN} International Conference on Functional Programming
  {(ICFP} '02), Pittsburgh, Pennsylvania, USA, October 4-6, 2002}},
  \bibfield{editor}{\bibinfo{person}{Mitchell Wand} {and}
  \bibinfo{person}{Simon L.~Peyton Jones}} (Eds.). \bibinfo{publisher}{{ACM}},
  \bibinfo{pages}{36--47}.
\newblock
\href{https://doi.org/10.1145/581478.581483}{doi:\nolinkurl{10.1145/581478.581483}}


\bibitem[Ford(2004a)]%
        {peg}
\bibfield{author}{\bibinfo{person}{Bryan Ford}.}
  \bibinfo{year}{2004}\natexlab{a}.
\newblock \showarticletitle{Parsing expression grammars: a recognition-based
  syntactic foundation}. In \bibinfo{booktitle}{\emph{Proceedings of the 31st
  ACM SIGPLAN-SIGACT Symposium on Principles of Programming Languages}}
  (Venice, Italy) \emph{(\bibinfo{series}{POPL '04})}.
  \bibinfo{publisher}{Association for Computing Machinery},
  \bibinfo{address}{New York, NY, USA}, \bibinfo{pages}{111–122}.
\newblock
\showISBNx{158113729X}
\href{https://doi.org/10.1145/964001.964011}{doi:\nolinkurl{10.1145/964001.964011}}


\bibitem[Ford(2004b)]%
        {ford04peg}
\bibfield{author}{\bibinfo{person}{Bryan Ford}.}
  \bibinfo{year}{2004}\natexlab{b}.
\newblock \showarticletitle{Parsing expression grammars: a recognition-based
  syntactic foundation}. In \bibinfo{booktitle}{\emph{Proceedings of the 31st
  {ACM} {SIGPLAN-SIGACT} Symposium on Principles of Programming Languages,
  {POPL} 2004, Venice, Italy, January 14-16, 2004}},
  \bibfield{editor}{\bibinfo{person}{Neil~D. Jones} {and}
  \bibinfo{person}{Xavier Leroy}} (Eds.). \bibinfo{publisher}{{ACM}},
  \bibinfo{pages}{111--122}.
\newblock
\href{https://doi.org/10.1145/964001.964011}{doi:\nolinkurl{10.1145/964001.964011}}


\bibitem[Fromherz and Protzenko(2024)]%
        {krml2rust}
\bibfield{author}{\bibinfo{person}{Aymeric Fromherz} {and}
  \bibinfo{person}{Jonathan Protzenko}.} \bibinfo{year}{2024}\natexlab{}.
\newblock \bibinfo{title}{Compiling C to Safe Rust, Formalized}.
\newblock
\showeprint[arxiv]{2412.15042}~[cs.PL]
\urldef\tempurl%
\url{https://arxiv.org/abs/2412.15042}
\showURL{%
\tempurl}


\bibitem[Hutton(1989)]%
        {Hutton89}
\bibfield{author}{\bibinfo{person}{Graham Hutton}.}
  \bibinfo{year}{1989}\natexlab{}.
\newblock \showarticletitle{Parsing Using Combinators}. In
  \bibinfo{booktitle}{\emph{Proceedings of the 1989 Glasgow Workshop on
  Functional Programming}}. \bibinfo{publisher}{Springer-Verlag},
  \bibinfo{address}{Berlin, Heidelberg}, \bibinfo{pages}{353–370}.
\newblock
\showISBNx{3540196099}


\bibitem[IEEE(2019)]%
        {ieee754}
\bibfield{author}{\bibinfo{person}{IEEE}.} \bibinfo{year}{2019}\natexlab{}.
\newblock \bibinfo{title}{IEEE 754-2019 - IEEE Standard for Floating-Point
  Arithmetic}.
\newblock \bibinfo{numpages}{84}~pages.
\newblock
\href{https://doi.org/10.1109/IEEESTD.2019.8766229}{doi:\nolinkurl{10.1109/IEEESTD.2019.8766229}}


\bibitem[Intel(2021)]%
        {tinycbor}
\bibfield{author}{\bibinfo{person}{Intel}.} \bibinfo{year}{2021}\natexlab{}.
\newblock \bibinfo{title}{TinyCBOR}.
\newblock \bibinfo{howpublished}{\url{https://github.com/intel/tinycbor}}.
\newblock


\bibitem[{ITU-T Study Group 17}(2021)]%
        {asn1}
\bibfield{author}{\bibinfo{person}{{ITU-T Study Group 17}}.}
  \bibinfo{year}{2021}\natexlab{}.
\newblock \bibinfo{title}{{X.680 : Information technology - Abstract Syntax
  Notation One (ASN.1): Specification of basic notation}}.
\newblock \bibinfo{howpublished}{ITU Recommendation X.680}.
\newblock
\urldef\tempurl%
\url{https://www.itu.int/rec/T-REC-X.680/}
\showURL{%
\tempurl}


\bibitem[Lasser et~al\mbox{.}(2019)]%
        {lasser19itp}
\bibfield{author}{\bibinfo{person}{Sam Lasser}, \bibinfo{person}{Chris
  Casinghino}, \bibinfo{person}{Kathleen Fisher}, {and} \bibinfo{person}{Cody
  Roux}.} \bibinfo{year}{2019}\natexlab{}.
\newblock \showarticletitle{A Verified {LL(1)} Parser Generator}. In
  \bibinfo{booktitle}{\emph{10th International Conference on Interactive
  Theorem Proving, {ITP} 2019, September 9-12, 2019, Portland, OR, {USA}}}
  \emph{(\bibinfo{series}{LIPIcs}, Vol.~\bibinfo{volume}{141})},
  \bibfield{editor}{\bibinfo{person}{John Harrison}, \bibinfo{person}{John
  O'Leary}, {and} \bibinfo{person}{Andrew Tolmach}} (Eds.).
  \bibinfo{publisher}{Schloss Dagstuhl - Leibniz-Zentrum f{\"{u}}r Informatik},
  \bibinfo{pages}{24:1--24:18}.
\newblock
\href{https://doi.org/10.4230/LIPIcs.ITP.2019.24}{doi:\nolinkurl{10.4230/LIPIcs.ITP.2019.24}}


\bibitem[Lasser et~al\mbox{.}(2021)]%
        {costar21}
\bibfield{author}{\bibinfo{person}{Sam Lasser}, \bibinfo{person}{Chris
  Casinghino}, \bibinfo{person}{Kathleen Fisher}, {and} \bibinfo{person}{Cody
  Roux}.} \bibinfo{year}{2021}\natexlab{}.
\newblock \showarticletitle{CoStar: a verified ALL(*) parser}. In
  \bibinfo{booktitle}{\emph{{PLDI} '21: 42nd {ACM} {SIGPLAN} International
  Conference on Programming Language Design and Implementation, Virtual Event,
  Canada, June 20-25, 20211}}, \bibfield{editor}{\bibinfo{person}{Stephen~N.
  Freund} {and} \bibinfo{person}{Eran Yahav}} (Eds.).
  \bibinfo{publisher}{{ACM}}, \bibinfo{pages}{420--434}.
\newblock
\href{https://doi.org/10.1145/3453483.3454053}{doi:\nolinkurl{10.1145/3453483.3454053}}


\bibitem[Lattuada et~al\mbox{.}(2024)]%
        {verus24sosp}
\bibfield{author}{\bibinfo{person}{Andrea Lattuada}, \bibinfo{person}{Travis
  Hance}, \bibinfo{person}{Jay Bosamiya}, \bibinfo{person}{Matthias Brun},
  \bibinfo{person}{Chanhee Cho}, \bibinfo{person}{Hayley LeBlanc},
  \bibinfo{person}{Pranav Srinivasan}, \bibinfo{person}{Reto Achermann},
  \bibinfo{person}{Tej Chajed}, \bibinfo{person}{Chris Hawblitzel},
  \bibinfo{person}{Jon Howell}, \bibinfo{person}{Jacob~R. Lorch},
  \bibinfo{person}{Oded Padon}, {and} \bibinfo{person}{Bryan Parno}.}
  \bibinfo{year}{2024}\natexlab{}.
\newblock \showarticletitle{Verus: A Practical Foundation for Systems
  Verification}. In \bibinfo{booktitle}{\emph{Proceedings of the ACM SIGOPS
  30th Symposium on Operating Systems Principles}} (Austin, TX, USA)
  \emph{(\bibinfo{series}{SOSP '24})}. \bibinfo{publisher}{Association for
  Computing Machinery}, \bibinfo{address}{New York, NY, USA},
  \bibinfo{pages}{438–454}.
\newblock
\showISBNx{9798400712517}
\href{https://doi.org/10.1145/3694715.3695952}{doi:\nolinkurl{10.1145/3694715.3695952}}


\bibitem[Lundblade(2023)]%
        {qcbor}
\bibfield{author}{\bibinfo{person}{Laurence Lundblade}.}
  \bibinfo{year}{2020--2023}\natexlab{}.
\newblock \bibinfo{title}{{QCBOR}}.
\newblock
  \bibinfo{howpublished}{\url{https://github.com/laurencelundblade/QCBOR}}.
\newblock


\bibitem[MITRE(2016)]%
        {pyrsa}
\bibfield{author}{\bibinfo{person}{MITRE}.} \bibinfo{year}{2016}\natexlab{}.
\newblock \bibinfo{title}{CVE-2016-1494}.
\newblock
  \bibinfo{howpublished}{\url{https://www.cve.org/CVERecord?id=CVE-2016-1494}}.
\newblock


\bibitem[Mundkur et~al\mbox{.}(2020)]%
        {parsley20}
\bibfield{author}{\bibinfo{person}{Prashanth Mundkur}, \bibinfo{person}{Linda
  Briesemeister}, \bibinfo{person}{Natarajan Shankar},
  \bibinfo{person}{Prashant Anantharaman}, \bibinfo{person}{Sameed Ali},
  \bibinfo{person}{Zephyr Lucas}, {and} \bibinfo{person}{Sean Smith}.}
  \bibinfo{year}{2020}\natexlab{}.
\newblock \showarticletitle{Research Report: The Parsley Data Format Definition
  Language}. In \bibinfo{booktitle}{\emph{2020 IEEE Security and Privacy
  Workshops (SPW)}}. \bibinfo{pages}{300--307}.
\newblock
\href{https://doi.org/10.1109/SPW50608.2020.00064}{doi:\nolinkurl{10.1109/SPW50608.2020.00064}}


\bibitem[Ni et~al\mbox{.}(2023)]%
        {asn1star}
\bibfield{author}{\bibinfo{person}{Haobin Ni}, \bibinfo{person}{Antoine
  Delignat-Lavaud}, \bibinfo{person}{C\'{e}dric Fournet},
  \bibinfo{person}{Tahina Ramananandro}, {and} \bibinfo{person}{Nikhil Swamy}.}
  \bibinfo{year}{2023}\natexlab{}.
\newblock \showarticletitle{ASN1*: Provably Correct, Non-malleable Parsing for
  ASN.1 DER}. In \bibinfo{booktitle}{\emph{Proceedings of the 12th ACM SIGPLAN
  International Conference on Certified Programs and Proofs}} (Boston, MA, USA)
  \emph{(\bibinfo{series}{CPP 2023})}. \bibinfo{publisher}{Association for
  Computing Machinery}, \bibinfo{address}{New York, NY, USA},
  \bibinfo{pages}{275–289}.
\newblock
\showISBNx{9798400700262}
\href{https://doi.org/10.1145/3573105.3575684}{doi:\nolinkurl{10.1145/3573105.3575684}}


\bibitem[Protzenko et~al\mbox{.}(2017)]%
        {lowstar}
\bibfield{author}{\bibinfo{person}{Jonathan Protzenko},
  \bibinfo{person}{Jean-Karim Zinzindohou\'{e}}, \bibinfo{person}{Aseem
  Rastogi}, \bibinfo{person}{Tahina Ramananandro}, \bibinfo{person}{Peng Wang},
  \bibinfo{person}{Santiago Zanella-B\'{e}guelin}, \bibinfo{person}{Antoine
  Delignat-Lavaud}, \bibinfo{person}{C\u{a}t\u{a}lin Hri\c{t}cu},
  \bibinfo{person}{Karthikeyan Bhargavan}, \bibinfo{person}{C\'{e}dric
  Fournet}, {and} \bibinfo{person}{Nikhil Swamy}.}
  \bibinfo{year}{2017}\natexlab{}.
\newblock \showarticletitle{Verified low-level programming embedded in F*}.
\newblock \bibinfo{journal}{\emph{Proc. ACM Program. Lang.}}
  \bibinfo{volume}{1}, \bibinfo{number}{ICFP}, Article \bibinfo{articleno}{17}
  (\bibinfo{date}{Aug.} \bibinfo{year}{2017}), \bibinfo{numpages}{29}~pages.
\newblock
\href{https://doi.org/10.1145/3110261}{doi:\nolinkurl{10.1145/3110261}}


\bibitem[Ramananandro et~al\mbox{.}(2019)]%
        {everparse}
\bibfield{author}{\bibinfo{person}{Tahina Ramananandro},
  \bibinfo{person}{Antoine Delignat-Lavaud}, \bibinfo{person}{Cedric Fournet},
  \bibinfo{person}{Nikhil Swamy}, \bibinfo{person}{Tej Chajed},
  \bibinfo{person}{Nadim Kobeissi}, {and} \bibinfo{person}{Jonathan
  Protzenko}.} \bibinfo{year}{2019}\natexlab{}.
\newblock \showarticletitle{{EverParse}: Verified Secure {Zero-Copy} Parsers
  for Authenticated Message Formats}. In \bibinfo{booktitle}{\emph{28th USENIX
  Security Symposium (USENIX Security 19)}}. \bibinfo{publisher}{USENIX
  Association}, \bibinfo{address}{Santa Clara, CA},
  \bibinfo{pages}{1465--1482}.
\newblock
\showISBNx{978-1-939133-06-9}
\urldef\tempurl%
\url{https://www.usenix.org/conference/usenixsecurity19/presentation/delignat-lavaud}
\showURL{%
\tempurl}


\bibitem[Reynolds(2002)]%
        {seplogic}
\bibfield{author}{\bibinfo{person}{John~C. Reynolds}.}
  \bibinfo{year}{2002}\natexlab{}.
\newblock \showarticletitle{Separation Logic: A Logic for Shared Mutable Data
  Structures}. In \bibinfo{booktitle}{\emph{Proceedings of the 17th Annual IEEE
  Symposium on Logic in Computer Science}} \emph{(\bibinfo{series}{LICS '02})}.
  \bibinfo{publisher}{IEEE Computer Society}, \bibinfo{address}{USA},
  \bibinfo{pages}{55–74}.
\newblock
\showISBNx{0769514839}


\bibitem[Schaad(2022)]%
        {cose}
\bibfield{author}{\bibinfo{person}{Jim Schaad}.}
  \bibinfo{year}{2022}\natexlab{}.
\newblock \bibinfo{title}{{CBOR Object Signing and Encryption (COSE):
  Structures and Process}}.
\newblock \bibinfo{howpublished}{IETF RFC 9052}.
\newblock
\href{https://doi.org/10.17487/RFC9052}{doi:\nolinkurl{10.17487/RFC9052}}


\bibitem[Shankar(1996)]%
        {pvs96}
\bibfield{author}{\bibinfo{person}{Natarajan Shankar}.}
  \bibinfo{year}{1996}\natexlab{}.
\newblock \showarticletitle{{PVS:} Combining Specification, Proof Checking, and
  Model Checking}. In \bibinfo{booktitle}{\emph{Formal Methods in
  Computer-Aided Design, First International Conference, {FMCAD} '96, Palo
  Alto, California, USA, November 6-8, 1996, Proceedings}}
  \emph{(\bibinfo{series}{Lecture Notes in Computer Science},
  Vol.~\bibinfo{volume}{1166})}, \bibfield{editor}{\bibinfo{person}{Mandayam~K.
  Srivas} {and} \bibinfo{person}{Albert~John Camilleri}} (Eds.).
  \bibinfo{publisher}{Springer}, \bibinfo{pages}{257--264}.
\newblock
\href{https://doi.org/10.1007/BFb0031813}{doi:\nolinkurl{10.1007/BFb0031813}}


\bibitem[Swamy et~al\mbox{.}(2016)]%
        {fstar}
\bibfield{author}{\bibinfo{person}{Nikhil Swamy},
  \bibinfo{person}{C\u{a}t\u{a}lin Hri\c{t}cu}, \bibinfo{person}{Chantal
  Keller}, \bibinfo{person}{Aseem Rastogi}, \bibinfo{person}{Antoine
  Delignat-Lavaud}, \bibinfo{person}{Simon Forest},
  \bibinfo{person}{Karthikeyan Bhargavan}, \bibinfo{person}{C\'{e}dric
  Fournet}, \bibinfo{person}{Pierre-Yves Strub}, \bibinfo{person}{Markulf
  Kohlweiss}, \bibinfo{person}{Jean-Karim Zinzindohou\'e}, {and}
  \bibinfo{person}{Santiago {Zanella-B\'eguelin}}.}
  \bibinfo{year}{2016}\natexlab{}.
\newblock \showarticletitle{Dependent Types and Multi-Monadic Effects in {F*}}.
  In \bibinfo{booktitle}{\emph{43rd ACM SIGPLAN-SIGACT Symposium on Principles
  of Programming Languages (POPL)}}. \bibinfo{publisher}{ACM},
  \bibinfo{pages}{256--270}.
\newblock
\showISBNx{978-1-4503-3549-2}
\urldef\tempurl%
\url{https://www.fstar-lang.org/papers/mumon/}
\showURL{%
\tempurl}


\bibitem[Swamy et~al\mbox{.}(2022)]%
        {everparse3d}
\bibfield{author}{\bibinfo{person}{Nikhil Swamy}, \bibinfo{person}{Tahina
  Ramananandro}, \bibinfo{person}{Aseem Rastogi}, \bibinfo{person}{Irina
  Spiridonova}, \bibinfo{person}{Haobin Ni}, \bibinfo{person}{Dmitry Malloy},
  \bibinfo{person}{Juan Vazquez}, \bibinfo{person}{Michael Tang},
  \bibinfo{person}{Omar Cardona}, {and} \bibinfo{person}{Arti Gupta}.}
  \bibinfo{year}{2022}\natexlab{}.
\newblock \showarticletitle{Hardening attack surfaces with formally proven
  binary format parsers}. In \bibinfo{booktitle}{\emph{Proceedings of the 43rd
  ACM SIGPLAN International Conference on Programming Language Design and
  Implementation}} (San Diego, CA, USA) \emph{(\bibinfo{series}{PLDI 2022})}.
  \bibinfo{publisher}{Association for Computing Machinery},
  \bibinfo{address}{New York, NY, USA}, \bibinfo{pages}{31–45}.
\newblock
\showISBNx{9781450392655}
\href{https://doi.org/10.1145/3519939.3523708}{doi:\nolinkurl{10.1145/3519939.3523708}}


\bibitem[{The FIDO Alliance}(2025)]%
        {ctap2-2}
\bibfield{author}{\bibinfo{person}{{The FIDO Alliance}}.}
  \bibinfo{year}{2025}\natexlab{}.
\newblock \bibinfo{title}{Client to Authenticator Protocol (CTAP), version
  2.2}.
\newblock
  \bibinfo{howpublished}{\url{https://fidoalliance.org/specs/fido-v2.2-ps-20250228/fido-client-to-authenticator-protocol-v2.2-ps-20250228.html}}.
\newblock


\bibitem[{Trusted Computing Group}(2023)]%
        {dpe}
\bibfield{author}{\bibinfo{person}{{Trusted Computing Group}}.}
  \bibinfo{year}{2023}\natexlab{}.
\newblock \bibinfo{title}{DICE Protection Environment, Version 1.0, Revision
  0.6}.
\newblock
  \bibinfo{howpublished}{\url{https://trustedcomputinggroup.org/wp-content/uploads/TCG-DICE-Protection-Environment-Specification_14february2023-1.pdf}}.
\newblock


\bibitem[{Word Wide Web Consortium}(2019)]%
        {webauthn}
\bibfield{author}{\bibinfo{person}{{Word Wide Web Consortium}}.}
  \bibinfo{year}{2019}\natexlab{}.
\newblock \bibinfo{title}{WebAuthn: An API for accessing Public Key
  Credentials}.
\newblock \bibinfo{howpublished}{\url{https://w3c.github.io/webauthn/}}.
\newblock


\bibitem[Zinzindohou\'e et~al\mbox{.}(2017)]%
        {haclstar}
\bibfield{author}{\bibinfo{person}{Jean-Karim Zinzindohou\'e},
  \bibinfo{person}{Karthikeyan Bhargavan}, \bibinfo{person}{Jonathan
  Protzenko}, {and} \bibinfo{person}{Benjamin Beurdouche}.}
  \bibinfo{year}{2017}\natexlab{}.
\newblock \showarticletitle{{HACL}*: {A} Verified Modern Cryptographic
  Library}. In \bibinfo{booktitle}{\emph{{ACM} Conference on Computer and
  Communications Security}}. \bibinfo{publisher}{ACM},
  \bibinfo{pages}{1789--1806}.
\newblock
\urldef\tempurl%
\url{http://eprint.iacr.org/2017/536}
\showURL{%
\tempurl}


\end{thebibliography}

\appendix

\section{A Recursive Format: Variable Arity Trees}
\label{appendix:arith}

Consider for instance a small integer arithmetic language with numeric
values, binary subtraction, and variable-arity addition. We specify
this language as a high-level \fstar inductive datatype:
\begin{lstlisting}[language=fstar]
type expr = | Value of U64.t | Minus of (expr * expr)
| Plus: (n: U8.t {n<254}) -> (l: nlist n expr) -> expr
\end{lstlisting}
We bound the number of addition operands to 253 because we want to
represent the node in the first byte: 255 for a value, followed by 8
bytes for the integer value; 254 for a subtraction, followed by 2
recursive payloads; otherwise, the object is an addition and the value
of the first byte gives the number of recursive operand payloads.

To this end, we specify a header parser for elements using
parser specification combinators:
\begin{lstlisting}[language=fstar]
let header = dtuple2 U8.t (fun h -> if h = 255 then U64.t else unit)
let parse_header = parse_u8 `parse_dtuple2` (fun fb ->
  if h = 255 then parse_u64 else parse_empty)
\end{lstlisting}
Note that the header contains the non-recursive integer value for the
value case, but does not contain any recursive payload for the
subtraction and addition cases.

Then, we define a \fstar function taking a header and determining the
number of recursive payloads needed:
\begin{lstlisting}[language=fstar]
let count_payloads (h: header) = let (| fb, ob |) = h in
  if fb = 255 then 0 else if fb = 254 then 2 else fb
\end{lstlisting}

Then, we define a \fstar function to turn a header and a list of
recursive expression payloads into an expression:
\begin{lstlisting}[language=fstar]
let synth (h: header) (pl: nlist (count h) expr) : expr = match h, pl with
| (| 255, v |), _      -> Value v
| (| 254, _ |), [a; b] -> Minus (a, b)
| (| n  , _ |), pl     -> Plus n pl
\end{lstlisting}
Then, we can call the recursive parser combinator to obtain the parser
specification for our expression language; thus enjoying validation in
constant stack space.

Then, using the zero-copy reader combinators we defined
in \pulseparse, we implement a shallow parser performing case analysis
on an expression implementation into the following implementation
datatype, leaving recursive payloads unparsed:
\begin{lstlisting}[language=fstar]
type parsed_to = | PValue of U64.t | PMinus of byte_array * byte_array
  | PPlus: (n: U8.t) -> (pl: byte_array) -> parsed_to
\end{lstlisting}
This datatype extracts to C as a tagged union.

By contrast, since we assume applications to have full control of
their memory consumption, we allow them to build arbitrarily nested
expressions, potentially containing some unparsed data for some
operands; thus, we provide the following implementation datatype from
which to serialize:
\begin{lstlisting}[language=fstar]
type serialize_from = | SBase of parsed_to
| SMinus of ref serialize_from * ref serialize_from
| SPlus: (len: U8.t) -> (pl: narray len serialize_from)
\end{lstlisting}
This datatype extracts to C as a tagged union, with \lsf{ref}
and \lsf{narray} extracting as C pointer types.

Then, using the copy writer combinators we defined in \pulseparse, we
implement a recursive serializer from values of this datatype.

The implementation of this example takes 150 lines of
specification and 1000 lines of \pulseparse implementation, which
extract to around 800 lines of C code. No proof was necessary, since
correctness and non-malleability are obtained by construction by
virtue of typechecking the combinator calls.
The full example is provided \ifanon{in the supplementary material}\else{in EverParse\footnote{See footnote \ref{footnote:everparse}}}\fi.

\section{Constant-stack, arithmetically safe validation of raw CBOR bytes}
\label{appendix:cbor-validator}
Let \lsc{size_t validate_header(uint8_t *input, size_t len)} be a
validator for CBOR item headers, returning the (nonzero) number of
bytes consumed for a valid header, and $0$ otherwise. Let
\begin{lstlisting}[language=c]
size_t get_payload_count(uint8_t *input, size_t len, bool *err)
\end{lstlisting}
be a function that takes a valid byte representation of a header and
returns the number of expected CBOR items to validate in the payload;
but sets \verb+*err+ to \verb+true+ if the expected length is greater
than the length \verb+len+ of its input, to avoid any arithmetic
overflow. Then, since the format of CBOR headers has the prefix
property (the validity of a header does not change if any bytes are
appended to it), the C function in Figure~\ref{fig:cbor-c-validator} \verb+validate+ is a memory
safe, arithmetically safe, and functionally correct validator for CBOR
items, returning the size of the valid CBOR item found at the
beginning of the input buffer, or 0 if none:

\begin{figure}
\begin{lstlisting}[language=C]
size_t validate(uint8_t *input, size_t len) {
  size_t expected = 1;
  size_t pos = 0;
  while (expected > 0) {
    expected = expected - 1;
    size_t header_size = validate_header(input+pos, len-pos);
    if (header_size == 0) return 0;
    pos = pos + header_size;
    if (expected > len - pos) return 0;
      // each remaining CBOR item consumes at least 1 byte
    bool err = false;
    size_t payload_count =
      get_payload_count(input+pos, header_size, &err);
    if (err) return false;
    if (payload_count > len - pos - expected) return false;
    expected = expected + payload_count;
  }
  return pos;
}
\end{lstlisting}
\caption{C validator for raw CBOR bytes} \label{fig:cbor-c-validator}
\end{figure}

\section{DICE Protection Environment}
\label{appendix:dpe}

DICE Protection Environment (DPE)~\cite{dpe} is a standard for a family of
protocols to measure and cryptographically attest the integrity of the boot
sequence of hardware ranging from IoT devices to cloud machines. DPE
implementations support various \emph{profiles}, exposing different interfaces
and capabilities to clients.~\citet{pulsecore} provide a verified implementation
of DPE in Pulse, supporting only the simplest profile, where a DPE client is
expected to be executing in the same address space, sharing memory with the DPE
attestation service. A more common profile instead allows a client to be
dislocated from the DPE service, and for them to communicate over a transport
using CBOR messages specified in CDDL.

The CDDL used in the DPE specification are all in a style that enable extension.
For example, all messages are of the form \verb+{ l => t,+
\verb+*(uint => any) }+, which, as explained in \S\ref{sec:cddl}, is ambiguous.
So, we adapt the specifications to add cuts, e.g., rewriting them to
\verb+{ l:t, *(uint => any)}+. Once in this form, \evercddl proves the
specifications unambiguous and generates Pulse code to parse and serialize CBOR
formatted messages to and from typed data structures. In total there are four
messages to the parsed as input to the DPE service and two messages that it
serializes as output back to the client.

We adapt Ebner et al.'s DPE interface and add a layer on top of it that adds
CDDL message parsing and serialization, with proofs in Pulse, demonstrating that
the specifications yielded by \evercddl are precise enough to express full
functional correctness of application code manipulating CBOR messages in a given
CDDL schema. For instance, our top-level specification of the \lsf{sign} API is
shown below:
\begin{lstlisting}[language=fstar]
  { input |->(p) i ** out |-> _ } let ok = sign input out
  { if ok=Success then ( exists o sig tbs. input |->(p) i ** out |-> o **
        (is_tbs_bytes tbs i /\ is_signature sig tbs /\ is_serialized_sig o sig)
    ) else ... }
\end{lstlisting}
This triple states that with (fractional) ownership of an input buffer with
bytes \lsf{i} and full ownership of an \lsf{out} buffer, if \lsf{sign} returns
\lsf{Success}, then the input buffer is unchanged, the output buffer contains
\lsf{o}, where \lsf{o} is a serialized signature \lsf{sig} of the to-be-signed
bytes \lsf{tbs} from a well-formatted input buffer \lsf{i}. We also fully
specify three possible modes of failure.

\end{document}